\documentclass[aps,showpacs,nofootinbib,superscriptaddress,showkeys,twocolumn,floatfix]{revtex4}

\usepackage{epsfig}
\usepackage{graphicx}
\usepackage{dcolumn}

\begin{document}

\title{Determination of low energy parameters for NN-scattering at
       ${\bf N^4 LO }$ in all partial waves with $j\le 5$. 
       }\author{M. Pav\'on Valderrama}\email{mpavon@ugr.es}
       \affiliation{Departamento de F\'{\i}sica Moderna, Universidad
       de Granada, E-18071 Granada, Spain.}  \author{E. Ruiz
       Arriola}\email{earriola@ugr.es} \affiliation{Departamento de
       F\'{\i}sica Moderna, Universidad de Granada, E-18071 Granada,
       Spain.}  \date{\today}

\begin{abstract} 
\rule{0ex}{3ex} The Variable S-matrix approach offers a unique way to
extract low energy threshold parameters for a given NN potential. We
extract those parameters for the np system from the NijmII and Reid93
potentials, to all partial waves with total angular momentum $j \le 5
$, using the generalized effective range expansion
$$
({\bf f}^{sj})_{l,l'} k^l k^{l'} = -({{\bf a}^{sj}}^{-1})_{l,l'} +
\frac12 ({\bf r}^{sj})_{l,l'} k^2 + ({\bf v_2}^{sj})_{l,l'} k^4 + 
+ ({\bf v_3}^{sj})_{l,l'} k^6 + + ({\bf v_4}^{sj})_{l,l'} k^8 + \dots
- {\rm i} k^{l+l'+1}
$$ 
where $ {\bf f}^{sj} = \left({\bf S}^{sj} - {\bf 1} \right) / (2 {\rm
i} k) $ is the scattering amplitude and ${\bf S}^{sj}$ is the unitary
S-matrix in coupled channel space with total spin $s$ and total
angular momentum $j$. Our calculation includes all the relevant
contributions of the full amplitude to order ${\cal O} (k^8) $ in the
CM momentum.  We also discuss the validity of the generalized
effective range expansion in the region of analyticity $ k \le m_\pi
/2$.
\end{abstract}


 \maketitle



\def\u{{\bf u}} \def\U{{\bf U}} \def\S{{\bf S}} \def\h{{\bf h}}
\def\L{{\bf L}} \def\E{{\bf 1}} \def\j{{\bf j}} \def\y{{\bf y}}
\def\M{{\bf M}} 
\def\f{{\bf f}}

\section{Introduction}

The study of the NN interaction has played a dominant role in the
theory of nuclear structure~\cite{Bethe:fi}.  At low energies the
corresponding scattering phase shifts can be best parameterized in
terms of an effective range expansion (ERE)~\cite{bethe_49,Noyes:zd}. This
expansion can be suitably generalized for all partial waves and
coupled channels (for a review see e.g. Ref.~\cite{Badalian:xj} and
references therein.). The calculation of the low energy threshold
parameters is straightforward in potential scattering, and can be
extracted from the asymptotic form of the wave function computed order
by order in a low energy expansion, but unlike potentials the ERE
parameters are not subjected to off-shell ambiguities. Benchmark
descriptions of NN scattering data have been obtained by the phase
shift partial wave analysis (PWA) of the Nijmegen group carried out a
decade ago~\cite{Stoks:1993tb} and further parameterizations through
high quality potentials~\cite{Stoks:wp}. For channels involving
central waves low energy threshold parameters have been determined
routinely and for the high quality potentials this calculation has
been undertaken in Ref.~\cite{deSwart:1995ui}). However, a detailed
and systematic determination of these threshold parameters in non
central partial waves is, to our knowledge, still missing.

Almost parallel with the previous developments there has been in the
last decade a renewed interest on the NN interaction based on the
application of Lagrangian effective field theories (EFT) methods,
motivated by Weinberg's ideas~\cite{Weinberg:rz,Weinberg:um} and
pursued by many
others~\cite{Ordonez:1992xp,Ordonez:tn,Ordonez:1995rz,vanKolck:1998bw,
Park:1997kp,Kaplan:1996xu,Kaplan:1998tg,Kaplan:1998we,Gegelia:gn,
Gegelia:1999ja,Fleming:1999ee,Fleming:1999bs,Epelbaum:1998ka,Epelbaum:1998na,Epelbaum:1999dj,Beane:2001bc,
Entem:2001cg,Entem:2002sf,Entem:2003ft,Epelbaum:2003gr,Epelbaum:2003xx,
Epelbaum:2004fk,Oller:px,Nieves:2003uu,Valderrama:2003np,Valderrama:2004nb}
to attempt a realistic description of np scattering data. This
approach has open promising perspectives in the theory of nuclear
systems (for a recent review see e.g. Ref.~\cite{Bedaque:2002mn} and
references therein).  Actually, the reference database most modern EFT
calculations are confronted with are those of the Nijmegen
group~\cite{Stoks:wp,Stoks:1993tb}. Moreover, the determination of the
low energy constants (LEC's) appearing in the EFT Lagrangian depends
on the regularization scale and also on the short distance behaviour
of the long distance chiral potential due to One, Two and higher Pion
exchanges.  The results for the scattering matrix and in particular
for the low energy threshold parameters should, of course, be
independent on the renormalization prescription. In the absence of
explicit pion exchanges the EFT approach can be mapped into an
effective range expansion~\cite{vanKolck:1998bw}, and the LEC's can,
in principle, be deduced from the physical threshold parameters for a
given renormalization scale. When long distance chiral potentials
generated by successive pion exchanges are included the situation
becomes much more involved, particularly if those potentials cannot be
treated perturbatively. Recently, we have proposed a regularization
scheme~\cite{Valderrama:2003np,Valderrama:2004nb} where the LEC's can
also be deduced directly from the low energy threshold parameters
beyond perturbation theory within a variable S-matrix
approach~\cite{Calogero}. The implementation of such an approach
requires an explicit knowledge of the threshold parameters as
input. In practice, most EFT calculations fit their results to the
Nijmegen database~\cite{Stoks:wp,Stoks:1993tb} in a given energy
window in terms of the LEC's within a given regularization scheme and
at a given scale, and low energy threshold parameters could be
determined afterwards. Under these circumstances we think it of
interest to provide in any case these low energy threshold parameters
as deduced {\it directly} from the high quality
potentials~\cite{Stoks:wp,Stoks:1993tb} in all partial waves.

In this paper we determine the low energy parameters of the NijmII and
Reid93 potentials~\cite{Stoks:1993tb} for all partial waves with $ j
\le 5 $. To this end we use the variable S-matrix approach (for a
review with many references see e.g. Ref.~\cite{Calogero}. An
application of the variable phase approach in the context of EFT in
the singlet $^1S_0$ channel can be found in Ref.~\cite{Steele:1998zc})
where the S-matrix can be computed by solving a non-linear system of
first order differential equations representing the variation of the
S-matrix under a continuous switching on of the potential starting
from a trivial potential (see below for a sketchy derivation). By
implementing a low energy expansion one obtains a set of first order
coupled differential equations from which the low energy threshold
parameters can be obtained as asymptotic values at long distances for
any partial wave. The method is stable and whenever comparison can be
made reproduces previous known results for the ERE parameters in
central waves. Partial aspects of the present paper have already
appeared~\cite{Valderrama:2004nb} in connection with the $^1S_0$ and
$^3S_1-^3D_1$ channels and in the context of the renormalization of
the singular OPE potential.

The paper is organized as follows. In Sect.~\ref{sec:smatrix} we
introduce our notation for scattering amplitudes, partial wave
expansion M-matrix and scaled M-matrix, which enable to take the low
energy expansion in a particularly transparent and clean way. We also
discuss the interplay between low energy expansion, rotational
covariance and unitarity. In Sect.~\ref{sec:vsmatrix} we briefly
outline the derivation of the variable S-matrix for coupled channels,
following the method of adiabatically switching on the potential from
the origin. Translating this equation to the variable scaled M-matrix
we write down in Sect.~\ref{sec:v_threshold} the coupled differential
equations which determine as a final condition the low energy
threshold parameters. Our numerical results are presented in
Sect.~\ref{sec:numres}. In Sect.~\ref{sec:lecs} we show the deduced
threshold parameters for the NijmII and Reid93 potentials for all
partial waves with total angular momentum $j \le 5$. In
Sect.~\ref{sec:ere} we examine the validity of the expansion for these
partial waves. In Sect.~\ref{sec:bound} we study the poles and
residues of the scattering matrix, mainly in connection with the
$^1S_0$ virtual state and the $^3S_1-^3D_1$ bound state deuteron
within the coupled channel effective range expansion.  Finally, in
Sect.~\ref{sec:concl} we present some conclusions and outlook for
further work. In Appendix~\ref{sec:app3} we provide some {\it raison
d'\^etre} for the present calculation and show the difficulties we
have encountered when a direct fit to the database is attempted.  In
the Appendix~\ref{sec:app} we investigate further the virtual and
bound state properties in an expansion of the inverse scattering
length in the $s-$wave channel.

\section{S-matrix, phase shifts and threshold parameters}
\label{sec:smatrix} 

As it is well known (see e.g. Ref.~\cite{Gloeckle}), the requirement
of all relevant symmetries in the NN-interaction implies that for the
two nucleon system in a total spin $s$ and total isospin $t$ state,
the elastic scattering amplitude for the transition $ \hat k , m_s \to
\hat k' , m'_ s$ with fixed CM energy $ E= k^2 / 2 \mu_{np} $
reads~\footnote{We use the normalization for the differential cross
section $$ \frac{d \sigma (\hat k , m_s \to \hat k' , m'_s ) }
{d\Omega} = | \f^{st}_{m'_s m_s} (\hat k' , \hat k ) |^2 $$.}
\begin{eqnarray}
\f^{st}_{m'_s m_s} (\hat k' , \hat k ) &=& \sum_{j l l' }
C_{l,l'}^{sj} (\hat k', \hat k ) \left[ 1-(-)^{l+s+t}\right]
\f^{sj}_{l,l'}
\label{eq:pw}
\end{eqnarray} 
where
\begin{eqnarray}
C_{l,l'}^{sj} (\hat k', \hat k ) &=& \sum_m {\rm i}^{l-l'} C(l,s,j
|m-m_s , m_s , m ) Y_{l,m-m_s} (\hat k) \nonumber
 \\ &\times&   C(l',s,j
 | m-m'_s , m'_s , m ) Y_{l',m-m'_s} (\hat k')^*
\end{eqnarray}
and  $\f^{sj}_{l,l'} $ is the scattering amplitude for the
$^{2s+1}l_j $ state with a total angular momentum $j$, total spin $s$
and orbital angular momentum $l$ and is given by 
\begin{eqnarray}
\f^{sj}_{l,l'} = \frac1{2{\rm i} k} \left( {\bf S}^{sj}_{l,l'}
-\delta_{l,l'} \right)
\end{eqnarray} 
in terms of the symmetric and unitary S-matrix, $ S^{sj}_{l',l} =
S^{sj}_{l,l'} $, $ \sum_{l'} S^{sj}_{l,l'} {S^{sj}_{l'',l'}}^* =
\delta_{l l'} $. Due to parity conservation, for $j> 0 $ the states
with $l=j$ cannot couple to states with $l=j \pm 1$. Thus, for the
spin singlet state, $s=0$, one has $l=j$ and hence the state is
uncoupled
\begin{eqnarray}
S_{jj}^{0j} = e^{ 2 i \delta_{j}^{0j} } \, 
\end{eqnarray}
whereas for the spin triplet state $s=1$, one has the uncoupled $ l=j$
state
\begin{eqnarray}
S_{jj}^{1j} &=& e^{  2 i \delta_{j}^{1j} } \;,
\end{eqnarray}
and the two channel coupled states $l,l'=j \pm 1$ states which we use
Stapp-Ypsilantis-Metropolis (SYM or Nuclear bar)~\cite{stapp}
parameterization for
\begin{eqnarray}
S^{1j} &=& \left( \begin{array}{cc} S_{j-1 \,
j-1}^{1j} & S_{j-1 \, j+1}^{1j} \\ S_{j+1 \, j-1}^{1j} & S_{j+1 \,
j+1}^{1j}
\end{array} \right) \nonumber  \\ &=& \left( \begin{array}{cc} \cos{(2
\bar \epsilon_j)} e^{2 i \bar \delta^{1j}_{j-1}} & i \sin{(2 \bar
\epsilon_j)} e^{i (\bar \delta^{1j}_{j-1} +\bar \delta^{1j}_{j+1})} \\
i \sin{(2 \bar \epsilon_j)} e^{i (\bar \delta^{1j}_{j-1} + \bar
\delta^{1j}_{j+1})} & \cos{(2 \bar \epsilon_j)} e^{2 i \bar
\delta^{1j}_{j+1}} \end{array} \right) \nonumber
\end{eqnarray}
In the discussion of low energy properties it is also interesting to
use the Blatt-Biedenharn (BB or Eigen phase)
parameterization~\cite{Bl52} that is given by
\begin{eqnarray}
S^{1j} &=& \left( \matrix{ \cos \epsilon_j & -\sin \epsilon_j \cr \sin
\epsilon_j & \cos \epsilon_j } \right) \left( \matrix{ e^{2 {\rm i}
\delta^{1j}_{j-1}} & 0 \cr 0 & e^{2 {\rm i} \delta_{j+1}^{1j}} }
\right) \nonumber \\ &\times& \left( \matrix{ \cos \epsilon_j & \sin
\epsilon_j \cr -\sin \epsilon_j & \cos \epsilon_j } \right)
\label{eq:BB} 
\end{eqnarray} 
The relation between the BB and SYM phase shifts is given by 
\begin{eqnarray}
\bar \delta_{j+1}^{1j} + \bar \delta_{j-1}^{1j} &=& \delta_{j+1}^{1j}
+ \delta_{j-1}^{1j} \, , \\ \sin( \bar \delta_{j-1}^{1j} - \bar
\delta_{j+1}^{1j}) &=& \frac{\tan( 2\bar \epsilon_j
)}{\tan(2\epsilon_j )} \, .
\end{eqnarray} 
In order to take the low energy limit we define the standard
symmetric and real coupled channel scaled M-matrix
\begin{eqnarray} 
\S = \left({\bf \hat M} + {\rm i} k {\bf D}^2 \right) \left({\bf \hat M} -
{\rm i} k {\bf D}^2 \right)^{-1} \, ,
\label{eq:M-matrix} 
\end{eqnarray} 
with $k$ the CM np momentum and ${\bf D} = {\rm diag} ( k^{l_1}, \dots , k^{l_N}) $.  
Due to unitarity of the S-matrix in the low energy limit, $ k\to 0$ we
have
\begin{eqnarray}
\left(\S - \E \right)_{l',l}=- 2 {\rm i} \alpha_{l', l}  k^{l'+l+1} +
\dots   \, ,  
\end{eqnarray} 
with $\alpha_{l' l} $ the (hermitian) scattering length matrix. The
threshold behaviour acquires its simplest form in the SYM
representation, 
\begin{eqnarray} 
\delta^{0j}_{j} &\to&  - \alpha^{0j}_{j} k^{2j+1} \, , \\ 
\delta^{1j}_{j} &\to&  - \alpha^{1j}_{j} k^{2j+1} \, , \\ 
\bar \delta^{1j}_{j-1} &\to&  - \bar \alpha^{1j}_{j-1} k^{2j-1} \, , \\ 
\bar \delta^{1j}_{j+1} &\to&  - \bar \alpha^{1j}_{j+1} k^{2j+3} \, , \\ 
\bar \epsilon_j &\to&  - \bar \alpha^{1j}_{j} k^{2j+1} \, . 
\label{eq:phase-thres}
\end{eqnarray} 
defining a natural hierarchy for phase-shifts (see also
below)~\footnote{In the BB form one has similar behaviours for the
$\delta$'s but $ \epsilon_j \to - \alpha^{1j}_j k^2 $.}.  The scaled
M-matrix, ${\bf \hat M} $, has a good low energy
behaviour~\cite{Badalian:xj} and admits the coupled channel analog of
the effective range expansion
\begin{eqnarray}
{\bf \hat M} = -{\bf a}^{-1} + \frac12 {\bf r}k^2 + {\bf v_2} k^4 +
{\bf v_3} k^6 +{\bf v_4} k^8 +
\dots \, , 
\label{eq:c-ere}
\end{eqnarray} 
where ${\bf a} $, ${\bf r}$ and ${\bf v_2} $ are the scattering length
matrix, effective range and curvature parameters respectively. Higher
order matrix parameters will be denoted for notational simplicity
${\bf v_3} $, ${\bf v_4}$, and so on. Within the ERE our notation is
as follows: LO means including order $k^0$ terms, NLO including $k^2$
terms and NNLO including $k^4$ terms, and so on. The expansion in $k^2
$ around the origin holds up to the next singularity, which in the
case of NN interaction corresponds to the pion left cut, so $ |k| \le
\pm m_\pi /2 \sim 70{\rm MeV} $. One of the advantages of the low
energy expansion at the level of the scaled M-matrix is that unitarity
is preserved exactly at any order in the expansion.

If one diagonalizes the M-matrix using the orthogonal transformation,
which we call $\Omega$, in Eq.~(\ref{eq:BB})
\begin{eqnarray}
\hat {\bf m} = 
{\bf D} \Omega {\bf D}^{-1} \, \hat{\bf M} \, {\bf D} \Omega^{-1} {\bf D} 
\end{eqnarray} 
the corresponding eigenvalues
are related to the BB eigen phaseshifts as follows
\begin{eqnarray}
\hat {\bf m}_l = k^{2l+1} \cot \delta_l^{sj} = -\frac1{\alpha_l^{sj}}
+ \frac12 r_l^{sj} k^2 + v_l^{sj} k^4 + \dots \, ,
\label{eq:c-ere-bb}
\end{eqnarray} 
for $l=j \pm 1$.

In connection with the low energy expansion note that the total
amplitude~(\ref{eq:pw}) satisfies both the off-forward optical theorem
from the unitarity of the S-matrix ($d \hat q $ represents solid angle
interaction in the unit vector $\hat q$ direction),
\begin{eqnarray}
\f^{st}_{m'_s m_s} (\hat k',\hat k) - \f^{st}_{m_s m'_s} (\hat k ,
\hat k' )^* &=&  \nonumber \\  \frac{{\rm i } k}{2\pi} \sum_{m_s''}
\int \, 
 d \hat q \, \f^{st}_{m''_s m_s} (\hat k' , \hat q ) \, \f^{st}_{m''_s
m_s} (\hat k , \hat q )^*  
\label{eq:opt_th}
\end{eqnarray} 
as well as rotational covariance, i.e. under rotations $ \vec k \to R
\vec k$ one has~\footnote{Corresponding to the invariance of $\bf f
(\vec k' , \vec k ) $.}
\begin{eqnarray}
\f^{st}_{m'_s m_s} (\hat k' , \hat k ) &\to& \sum_{m''_s,m'''_s}
D^{s}_{m'_s,m''_s} (R) \nonumber \\ &\times & 
\f^{st}_{m'_s m_s} ( R \hat k' , R \hat k ) D^{s}_{m'''_s,m_s} (R)^* 
\nonumber \\ 
\label{eq:rot_cov} 
\end{eqnarray} 
with $ D^{s}_{m'_s,m''_s} (R) $ the rotation Wigner matrices
corresponding to spin $ s$. This induces a mixing of the coupled
channel states, since they are irreducible under rotations. The latter
holds for any truncated sum in the partial wave expansion in
Eq.~(\ref{eq:pw}). On the other hand, the low energy expansion of the
amplitude only satisfies this properties perturbatively; the
coefficients of $\vec k$ and $\vec k'$ in a Taylor expansion of the
amplitude $\f^{st}_{m'_s m_s} (\hat k' , \hat k )$ are neither
rotational covariance nor do they satisfy exact unitarity. This can be
seen from the coupled channel amplitude,
\begin{eqnarray}
{\bf f} = {\bf D} \left[ \hat \M - i k {\bf D}^2 \right]^{-1} {\bf D} 
\end{eqnarray} 
which yields after taking the $ k\to  0$ expansion yields  
\begin{eqnarray}
{\bf f} = - {\bf D} {\bf a} {\bf D} - \frac12 {\bf D} {\bf a} {\bf r}
{\bf a} {\bf D}  k^2 + \dots 
\end{eqnarray} 
This threshold behaviour sets up a natural low energy hierarchy for
the $^{2s+1} l_j $ states
\begin{eqnarray} 
{\cal O} ( k^0 ) && ^1S_0 \, , ^3S_1 \nonumber \\ 
{\cal O} ( k^2 ) && ^1P_1 \, , ^3P_0 \, , ^3P_1 \, , ^3P_2 , E_1
\nonumber \\ 
{\cal O} ( k^4 ) && ^1D_1 \, , ^3D_1 \, , ^3D_2 \, , ^3D_3 , E_2
\nonumber \\ 
{\cal O} ( k^6 ) && ^1F_3 \, , ^3F_2 \, , ^3F_3 \, , ^3F_4 , E_3
\nonumber \\ 
{\cal O} ( k^8 ) && ^1G_4 \, , ^3G_3 \, , ^3G_4 \, , ^3G_5 , E_4
\end{eqnarray} 
where we use the notation $ E_1 = ^3S_1-^3D_1$ , $ E_2 = ^3P_2-^3F_2$
, $ E_3 = ^3D_3-^3G_3$ , $ E_4 = ^3F_4-^3H_4$, and so on. (in general
$E_j= ^3 (j+1)_j - ^3(j-1)_j )$. This means that for a complete
calculation at ${\cal O} (k^8) $ we have to include the $\alpha$'s for
the states $ ^1G_4 \, , ^3G_3 \, , ^3G_4 \, , ^3G_5 $ and$ E_4 $,
$\alpha$'s and $r$'s for $ ^1F_3 \, , ^3F_2 \, , ^3F_3 \, , ^3F_4 $
and $ E_3 $, $\alpha$'s, $r$'s and $v_2$'s for $ ^1D_1 \, , ^3D_1 \, ,
^3D_2 \, , ^3D_3 $ and $ E_2 $ and $\alpha$'s, $r$'s $v_2$'s and $v_3$'s
for $ ^1P_1 \, , ^3P_0 \, , ^3P_1 \, , ^3P_2 , E_1 $, and $\alpha$'s,
$r$'s, $v_2$'s, $v_3$'s and $v_4$'s for $ ^1S_0 $ and $^3S_1$. In general,
to ${\cal O} (k^{2n} )$ one needs
\begin{eqnarray}
N=2 \times (n+1) + 5\times \frac12 n (n+1)  
\end{eqnarray}
independent parameters. Thus, at ${\cal O} (k^0) $ we need only $N=2$
parameters, at ${\cal O} (k^2) $ one has $ N= 9$ parameters, at ${\cal
O} (k^4) $ , $ N= 21 $, at ${\cal O} (k^6) $ , $ N= 38 $ and at ${\cal
O} (k^8) $ we have $ N=60$ parameters, etc. Although these seem rather
large numbers, note that the declared number of independent adjustable
parameters to NN data in the high quality potential of
Ref.~\cite{Stoks:wp} is 41. The very recent EFT calculation NNNLO in
the Weinberg counting~\cite{Epelbaum:2004fk} declares 26 adjustable
parameters (actually LEC's) to the NN database of
Ref.~\cite{Stoks:wp}. This calculation corresponds to go to order $
{\cal O} (k^6) $ for the theory without explicit pions, i.e. for $ k
\ll m_\pi /2 $.

In this paper we use the low energy expansion in a way that both
unitarity~(\ref{eq:opt_th}) and rotational
covariance~(\ref{eq:rot_cov}) are simultaneously satisfied. As a
consequence, the low energy expansion for the amplitude is not
complete in the sense that starting at a given order on $k$ not all
the contributions of higher order are taken into
account. Nevertheless, up to order ${\cal O} (k^8) $ we compute all
the 60 parameters.  The highest angular momentum involving this order
is $ j = 5 $.  For completeness we present all the effective range
parameters up to $ v_4 $ in all partial waves with $ j\le 5 $. We
will see below, however, that presenting the data in terms of the
scaled M-matrix provides some interesting insights regarding the
validity of the effective range expansion.

\section{ Variable S-matrix}
\label{sec:vsmatrix}

In this Section we re-derive the variable S-matrix equation using a
continuous deformation of the potential. Alternative derivations are
based on Jost functions~\cite{Calogero} and inner boundary
conditions~\cite{Valderrama:2003np,Valderrama:2004nb}. The coupled
channel Schr\"odinger equation for the relative motion reads
\begin{eqnarray}
-\u '' (r) + \left[ \U (r) + \frac{{\bf l}^2}{r^2} \right] \u (r) =
 k^2 \u (r) \, , 
\label{eq:sch_cp} 
\end{eqnarray} 
where $\U (r)= 2 \mu_{np} {\bf V}(r)$ is the coupled channel matrix
potential which can be written as for $j> 0$,
\begin{eqnarray}
\U^{0j} (r) &=& U_{jj}^{0j} \nonumber \\ \\   
\U^{1j} (r) &=& \left( \matrix{ U_{j-1,j-1}^{1j} (r) & 0 &
U_{j-1,j+1}^{1j} (r) \cr 0 & U_{jj}^{1j} (r) & 0 \cr U_{j-1,j+1}^{1j}
(r) & 0 & U_{j+1,j+1}^{1j} (r) } \right) \, \nonumber  
\end{eqnarray} 
We will take for these potentials the ones available in
Ref.~~\cite{Stoks:1993tb}. In Eq.~(\ref{eq:sch_cp}) $ {\bf l}^2 = {\rm
diag} ( l_1 (l_1+1), \dots, l_N (l_N +1) )$ is the angular momentum,
$\u(r)$ is the reduced matrix wave function and $k$ the
C.M. momentum. In the case at hand $N=1$ for the spin singlet channel
with $l=j$ and $ N=3 $ for the spin triplet channel with $l_1=j-1$,
$l_2=j$ and $l_3=j+1$. For ease of notation we will keep the compact
matrix notation of Eq.~(\ref{eq:sch_cp}). The potentials in
Ref.~\cite{Stoks:1993tb} are regular at the origin, so the regular
solution is given by the boundary condition at the origin
\begin{eqnarray}
\u (0) = 0 \, ,  
\label{eq:bc}
\end{eqnarray}   
At long distances, we assume the asymptotic normalization condition
\begin{eqnarray}
\u (r)  \to \hat \h^{(-)} (r) - \hat \h^{(+)} (r) \S 
\label{eq:asym}
\end{eqnarray} 
with $\S$ the standard coupled channel unitary S-matrix. The
corresponding out-going and in-going free spherical waves are given by
\begin{eqnarray}
\hat \h^{(\pm)} (r) &=& {\rm diag} ( \hat h^\pm_{l_1} ( k r) , \dots ,
\hat h^\pm_{l_N} (k r) ) \, ,
\end{eqnarray} 
with $ \hat h^{\pm}_l ( x) $ the reduced Hankel functions of order
$l$, $ \hat h_l^{\pm} (x) = x H_{l+1/2}^{\pm} (x) $ ( $ \hat h_0^{\pm}
= e^{ \pm i x}$ ), and satisfy the free Schr\"odinger's equation for a
free particle. 

In order to deduce a variable $S$-matrix equation, we determine first
the infinitesimal change of the $S$ matrix under a general deformation
of the potential $\U (r) \to \U(r) + \Delta \U(r) $. Using
Schr\"odinger's equation (\ref{eq:sch_cp}) and the standard Lagrange's
identity adapted to this particular case, we get 
\begin{eqnarray}
\left[ \u (r)^\dagger \Delta \u'(r) - \u'(r)^\dagger \Delta \u(r)
\right]' = 
\u(r)^\dagger \Delta \U(r) \u(r) \nonumber \\ 
\end{eqnarray} 
which, after integration from the origin to infinity and using the
asymptotic form of the matrix wave function, Eq.~(\ref{eq:asym}), as
well as the regular condition at the origin, Eq.~(\ref{eq:bc}) yields
\begin{eqnarray}
2 {\rm i} k \S^\dagger \Delta \S = \int_0^\infty dr \, \u(r)^\dagger
\Delta \U(r) \u(r)
\end{eqnarray} 
In particular, for the parametric family of potentials $\bar \U (r,R)
= \theta(R-r) \U(r) $ we get $ \Delta \bar \U (r,R)
= \delta(R-r) \U(r) \Delta R $ and hence 
\begin{eqnarray}
2 {\rm i} k \S^\dagger (R) \S '(R)  =  \u (R)^\dagger \U(R) \u(R) \, , 
\end{eqnarray} 
and using the value of the wave function at the outer boundary,
Eq.~(\ref{eq:asym}) 
\begin{eqnarray}
\u (R)  =  \h^{(-)} (R) -  \h^{(+)} (R)  \S (R)  \,  , 
\end{eqnarray} 
we finally get the variable S-matrix equation, 
\begin{eqnarray}
2 {\rm i} k \frac{ d \S (R)}{dR} &=& \left[ \S(R) \hat \h^{(+)} (R) - \hat \h^{(-)}
(R) \right] \U(R) \nonumber \\ & \times& \left[ \hat \h^{(-)} (R) - \hat \h^{(+)} (R) \S (R)
\right] \, . 
\label{eq:vs}
\end{eqnarray} 
This is a first order non-linear matrix differential equation which
can be solved by standard means, provided the S-matrix is known at one
given scale. Note that for any value of the boundary radius we have a
different on-shell scattering problem. In the case of a regular
potential, Eq.~(\ref{eq:vs}) has to be supplemented with an initial
condition at the origin, namely the trivial one (corresponding to the
absence of a potential), and its asymptotic value yields the full
$S-$matrix ;
\begin{eqnarray}
\S(0) = {\bf 1} \qquad , \qquad  \S  = \S (\infty)   
\label{eq:ini_reg} 
\end{eqnarray} 

\section{Evolution of low energy parameters}
\label{sec:v_threshold} 
In order to take the low energy limit of Eq.~(\ref{eq:vs}) and
corrections there-off, we introduce the variable or running
$M-$matrix, in analogy with Eq.~(\ref{eq:M-matrix})  
\begin{eqnarray}
\S (R) = \left( {\bf \hat M} (R)+ {\rm i} k {\bf D}^2 \right)
\left({\bf \hat M}(R) - {\rm i} k {\bf D}^2 \right)^{-1} \, ,
\end{eqnarray} 
as well as the reduced Bessel functions 
\begin{eqnarray}
{\bf \hat j} &=& \frac1{2 {\rm i}} \left( \hat \h^{(+)} - \hat
\h^{(-)} \right) \, , \\ -{\bf \hat y} &=& \frac12 \left( \hat \h^{(+)} +
\hat \h^{(-)} \right) \, , 
\end{eqnarray} 
with 
\begin{eqnarray}
\hat j_l (x) = x j_l (x) \, , \qquad \,  \hat y_l (x) = x y_l
(x) \, . 
\end{eqnarray} 
Thus, we get
\begin{eqnarray}
\hat \M ' (k,R) &=& \left( \hat \M (R, k) \frac1k \hat {\bf j} (kR)
{\bf D}^{-1} - \hat {\bf y} (kR) {\bf D} \right) \U (R) \nonumber \\ &
\times & \left( \frac1k \hat {\bf j} (kR) {\bf D}^{-1} \hat \M (R, k)
- \hat {\bf y} (kR) {\bf D} \right) \, .
\label{eq:vkhat} 
\end{eqnarray} 
The scaled variable M-matrix admits the analog of the effective range
expansion
\begin{eqnarray}
{\bf \hat M} (R) = -{\bf a} (R)^{-1} + \frac12 {\bf r} (R) k^2 + {\bf
v_2} (R) k^4 + {\bf v_3} (R) k^6 + \dots \, ,
\end{eqnarray} 
where ${\bf a} (R)$, ${\bf r}(R)$, ${\bf v_2}(R)$, ${\bf v_3}(R)$, ${\bf
v_4}(R)$ , .. are the corresponding running parameters.  In this form
the low energy limit can be easily taken. Defining the matrix
functions 
\begin{eqnarray}
{\bf A}_k (R) &=& {\rm diag} \left( \frac{\hat j_{l_1} (kR)}{k^{l_1+1}}
, \dots, \frac{\hat j_{l_N} (kR)}{k^{l_N+1}}\right) \nonumber \\ {\bf
B}_k(R) &=& {\rm diag} \left( \hat y_{l_1} (kR) k^{l_1} , \dots ,  \hat
y_{l_N} (kR)k^{l_N} \right)
\end{eqnarray}
and their low energy expansion 
\begin{eqnarray}
{\bf A}_k(R) &=& \frac{\hat {\bf j} (kR)}k {\bf D}^{-1} = {\bf A}_0 +
k^2 {\bf A}_2 + k^4 {\bf A}_4 + \dots \, , \nonumber \\ {\bf B}_k(R)
&=& \hat {\bf y} (kR) {\bf D} \quad = {\bf B}_0 + k^2 {\bf B}_2 + k^4
{\bf B}_4 + \dots \, , \nonumber \\
\end{eqnarray} 
we get the system of coupled equations
\begin{widetext}
\begin{eqnarray} 
\frac{d}{dR} [{\bf a} (R)]^{-1} &=& -\left( [{\bf a} (R)]^{-1} {\bf
A}_0 + {\bf B}_0 \right) \U (R) \left( {\bf
A}_0 [{\bf a} (R)]^{-1} + {\bf B}_0 \right) \, , \nonumber \\ 
 \frac{d}{dR} {\bf r} (R) &=& \left( [{\bf a} (R)]^{-1} {\bf A}_0 +
{\bf B}_0 \right) \U (R) \left( {\bf r} (R) {\bf A}_0 + 2 [{\bf a}
(R)]^{-1} {\bf A}_2 + 2 {\bf B}_2 \right) \nonumber \\ &+& \left( {\bf
r} (R) {\bf A}_0 + 2 [{\bf a} (R)]^{-1} {\bf A}_2 + 2 {\bf B}_2
\right) \U (R) \left( [{\bf a} (R)]^{-1} {\bf A}_0 + {\bf B}_0 \right)
\, ,  \label{eq:var}
\\
 \frac{d}{dR} {\bf v_2} (R) &=& \left( [{\bf a} (R)]^{-1} {\bf A}_0 +
{\bf B}_0 \right) \U (R) \left( - [{\bf a} (R)]^{-1} {\bf A}_4 +
\frac12 {\bf r} (R) {\bf A}_2 + {\bf v_2} (R) {\bf A}_0 -{\bf B}_4
\right) \nonumber \\ &+& \left( - [{\bf a} (R)]^{-1} {\bf A}_4 +
\frac12 {\bf r} (R) {\bf A}_2 + {\bf v_2} (R) {\bf A}_0 -{\bf B}_4
\right) \U(R) \left( [{\bf a} (R)]^{-1} {\bf A}_0 + {\bf B}_0 \right)
\nonumber \\ &+& \left( \frac12 {\bf r} (R) {\bf A}_2 - [{\bf a}
(R)]^{-1} {\bf A}_2 - {\bf B}_2 \right) \U (R) \left( \frac12 {\bf r}
(R) {\bf A}_2 - [{\bf a} (R)]^{-1} {\bf A}_2 - {\bf B}_2 \right) \,
, \nonumber 
\end{eqnarray} 
\end{widetext}
and similar equations for ${\bf v_3} (R) $ and ${\bf v_4} (R)$. These
equations generalize to the coupled channel case those already found
in Ref.~\cite{Valderrama:2003np} and have to be supplemented with the
initial conditions,
\begin{eqnarray} 
{\bf a} (0)= 0 \, , \qquad {\bf r} ( 0)= 0  , \qquad {\bf v_2} ( 0)= 0 
\qquad ,   \dots 
\label{eq:ini} 
\end{eqnarray} 
The physical threshold parameters correspond to the values at
infinite,   
\begin{eqnarray} 
{\bf a} ( \infty)= {\bf a} \, , \qquad {\bf r} ( \infty)= {\bf r} \, , \qquad {\bf v_2} ( \infty)= {\bf v_2}
\qquad , \dots
\end{eqnarray} 
In the triplet coupled channel case the SYM threshold parameters matrices
are
\begin{eqnarray}
{\bf a}^{1j} &=& \left(\matrix{ \bar \alpha_{j-1}^{1j} &
\bar \alpha_{j}^{1j} \cr \bar \alpha_{j}^{1j} &
\bar \alpha_{j+1}^{1j} } \right) \, , \\ {\bf r}^{1j} &=&
\left(\matrix{ \bar r_{j-1}^{1j} & \bar r_{j}^{1j} \cr
\bar r_{j}^{1j} & \bar r_{j+1}^{1j} } \right) \, , \\ {\bf v}^{1j}
&=& \left(\matrix{ \bar v_{j-1}^{1j} & \bar v_{j}^{1j} \cr
\bar v_{j}^{1j} & \bar v_{j+1}^{1j} } \right) \, , 
\end{eqnarray} 
and so on. Similar definitions hold for the uncoupled channels.

\section{Numerical Results} 
\label{sec:numres}
\subsection{Determination of low energy parameters}
\label{sec:lecs}

Low energy scattering data can directly be described in terms of
threshold parameters, like $\alpha$, $r$, etc. , defined through
Eq.~(\ref{eq:c-ere}). Unfortunately, besides $ \alpha $ and $r_0$ in
the $^1S_0$ and $^3S_1$ channels, the partial wave analysis data
base~\cite{Stoks:1993tb} does not provide values for them except for $
v_2,v_3,v_4$ in Ref.~\cite{deSwart:1995ui} in the deuteron channel. In
Appendix~\ref{sec:app3} we show that a direct fit to the database
turns out to be numerically unreliable. The NN data base provides
explicit potentials like the NijmII and Reid93 potentials, for which
the variable phase approach may directly be applied. In such a way we
can uniquely and accurately determine all the needed low energy
threshold parameters by integrating Eqs.~(\ref{eq:var}) from the
origin to infinity with the trivial initial conditions
Eq.~(\ref{eq:ini}). In all cases we have checked that the variable
S-matrix equation, Eq.~(\ref{eq:vs}), indeed reproduces the results of
the NN data base~\cite{Stoks:1993tb} for the phase shifts. In
practice, our results are stable if one integrates up to $R_\infty =
40 {\rm fm}$. Our results for the low energy threshold parameters for
the NijmII and Reid93 potentials can be summarized in
Table~\ref{tab:table1}. To obtain these results we looked for a stable
plateau in the large $R$ region in each threshold parameter
separately. So we quote in Table~\ref{tab:table1} the stable digits in
this flat region.  This is the main result of this work. In general we
see a rather good agreement between both potentials never worse than $
5\% $ and frequently much better. We also observe, as expected, that
numerical accuracy worsens by going to higher orders in the ERE in a
given partial wave or going to higher partial waves.

In the interesting case of the $^3S_1$ eigen-channel, one
has~\footnote{We use the notation $v=v_2$,$v'=v_3$ and $v''=v_4$ for
simplicity}
\begin{eqnarray}
k \cot \delta_{3S1} = -\frac1{\alpha_{3S1}}+ \frac12 r_{3S1}
k^2 + v_{3S1} k^4 + \dots
\label{eq:keigen}
\end{eqnarray}  
where we get for the effective range parameters the relations, 
\begin{eqnarray}
\alpha_{3S1} &=& \bar \alpha_{3S1} \nonumber \\ 
r_{3S1} &=& \bar
r_{3S1}+ \frac{2 \bar r_{E1} \bar \alpha_{E1}}{\bar \alpha_{3S1}}+
\frac{ \bar r_{3D1} \bar \alpha_{E1}^2}{\bar \alpha_{3S1}^2 } \\
v_{3S1} &=& \bar v_{3S1} + \frac14 \bar \alpha_{3D1} \bar r_{E1}^2
\nonumber \\ 
&+& \frac{\bar \alpha_{E1}}{4 \bar \alpha_{3S1}} \left (
2 \bar \alpha_{3D1} \bar r_{3D1} \bar r_{E1} - \bar \alpha_{E1} \bar
r_{E1}^2 + 8 \bar v_{E1} \right) \nonumber \\
 &+& \frac{\bar \alpha_{E1}^2}{4\bar \alpha_{3S1}^2} \left( 
\bar \alpha_{3D1} \bar r_{3D1}^2 - 2 
 \bar \alpha_{E1} \bar r_{3D1} \bar r_{E1} + 4 \bar v_{3D1} \right)
 \nonumber \\ 
&+& \frac1{4 \bar \alpha_{3S1}^3} \left( 4 \bar \alpha_{E1}^2 - \bar
 \alpha_{E1}^4 \bar r_{3D1}^2 \right)
\label{eq:sym-eigen} 
\end{eqnarray} 
and so on. Using the numerical values for the NijmII (Reid93)
potentials from Table~\ref{tab:table1} we get
\begin{eqnarray}
\alpha_{3S1} &=& 5.41896 \, (5.42293) \, {\rm fm} \nonumber \\ 
r_{3S1} &=& 1.75334 \, (1.75556) \, {\rm fm} \nonumber \\ 
v_{3S1} &=& 0.04531 \, (0.03266)\, {\rm fm}^3 \nonumber \\
v'_{3S1} &=& 0.65831 \, (0.65797) \, {\rm fm}^5 \nonumber \\
v''_{3S1} &=& -4.19144 \, (-4.19262) \, {\rm fm}^7 \nonumber \\
\end{eqnarray} 
in agreement with previous
findings~\cite{deSwart:1995ui,Epelbaum:2004fk}.


\begin{table*}
\caption{\label{tab:table1} Low energy threshold parameters for all
partial waves in np scattering for the NijmII and Reid93 (in brackets)
potentials. We give the scattering length $\alpha$, the effective
range $r$ and the curvature parameters $v_2$, $v_3$ and $v_4$.  Units
are in relevant powers of a fm, with $ l', l = j, j\pm 1 $. From them
one can compute the M-matrix at low energies using $ {\bf M}_{l,l'}
k^l k^{l'} = -({\bf a}^{-1})_{l,l'} + \frac12({\bf r})_{l,l'} k^2
+({\bf v_2})_{l,l'} k^4+({\bf v_3})_{l,l'} k^6+({\bf v_4})_{l,l'} k^8 +
\dots $ where $ {\bf M} = \left({\bf S} - 1 \right) \left({\bf S} +1
\right)^{-1} {\rm i} k $ and ${\bf S}$ is the unitary S-matrix in
coupled channel space. In the coupled channel case we use the
SYM(Nuclear bar) low energy parameters. All presented digits are
numerically significant. A long dash (---) stands for cases where
numerical accuracy was outraged and no value could be reliably
deduced.}
\begin{ruledtabular}
\begin{tabular}{|c|c|c|c|c|c|}
\hline 
Wave  & $ \alpha({\rm fm}^{l+l'+1}) $ & $r_0({\rm fm}^{l+l'+1})
$ & $v_2({\rm fm}^{l+l'+3}) $ & $v_3({\rm fm}^{l+l'+5}) $  & $v_4({\rm fm}^{l+l'+7}) $ \\ 
 & & & &  &  \\ 
 & NijmII \,( Reid93 )  & NijmII \,( Reid93 )   
 & NijmII \,( Reid93 )  & NijmII \,( Reid93 ) & NijmII\, ( Reid93 )  \\ 
\hline 
\hline $^1S_0 $ & -23.727(-23.735) & 2.670(2.753) & -0.4759(-0.4942) & 3.962(3.652) & -19.88(-18.30) \\
\hline $^3P_0 $ & -2.468(-2.469) & 3.914(3.870) & 1.099(0.9616) & 3.816(3.712) & -7.6(-7.4)\\
\hline
\hline 
$^1P_1 $ & 2.797(2.736) & -6.399(-6.606) & -1.580(-1.834) & 0.404(1.024) & 7.8(8.4) \\
\hline $^3P_1 $ & 1.529(1.530) & -8.580(-8.556) & -0.0181(+0.01006)& -0.86(-0.934) & 0.3(0.2) \\
\hline 
$^3S_1 $ & 5.418(5.422) & 1.833(1.833) & -0.131(-0.141) & 1.444(1.433) & -7.95(-7.9) \\
$^3D_1 $ & 6.505(6.453) & -3.523(-3.566) & -3.699(-3.803) & 1.12(1.023) & -3(-2.6) \\
$E_1 $ & 1.647(1.645) & 0.404(0.413) & -0.274(-0.264) & 1.447(1.423) & -7.3(-7.3)   \\
\hline 
\hline 
$^1D_2 $ & -1.389(-1.377) & 14.87(15.04) & 16.37(16.73) & -13.2(-13.0) & 34.0(34.0) \\
\hline $^3D_2 $ & -7.405(-7.411) & 2.858(2.851) & 2.395(2.370) & -0.99(-1.00) & 2.0(2.0) \\
\hline $^3P_2 $ & -0.2844(-0.2892) & -8.270(-8.363) & -6.91(-7.13) & -6.0(-6.3) & -20(-30)\\
$^3F_2 $ & -0.9763(-0.9698) & -5.640(-5.821) & -22.96(-23.78) & -79.32(-83.0) & -117(-127) \\
$E_2 $ & 1.609(1.600) & -15.70(-15.89) & -25.18(-25.72) & -23.3(-24.8) & -67(-70)\\
\hline
\hline 
$^1F_3 $ & 8.383(8.365) & -3.924(-3.936) & -9.888(-9.937) & -15.4(-15.6) & -4(-4)\\
\hline $^3F_3 $ & 2.703(2.686) & -9.932(-9.994) & -20.56
(-20.73) & -19(-19.4) & -20(-20)\\
\hline 
$^3D_3 $ & -0.1449(-0.1770) & 1.369(1.365) & 2.06(2.040) & 2(2) & ---(---) \\
$^3G_3 $ & 4.880(4.874) & -0.03306(-0.0529) & -0.0166(-0.1261) & -0.117(-0.662) & -3.1(-4.1) \\
$E_3 $ & -9.695(-9.683) & 3.255(3.249) & 7.655(7.618) & 9.5(9.3) & ---(---) \\
\hline 
\hline 
$^1G_4 $ & -3.229(-3.210) & 10.78(10.82) & 34.4(34.50) & 80(84) & ---(---)\\
\hline $^3G_4 $ & -19.17(-19.14) & 2.056(2.059) & 6.810
(6.822) & 16.6(16.7) & 10(14)\\
\hline 
$^3F_4 $ & -0.01045(-0.01053) & -3.02(-3.04) & -8.0(-8.0) 
& -10(---) & ---(---)\\
$^3H_4 $ & -1.250(-1.240) & -0.1762(-0.178) & -1.52(-1.527) & -11(-11) & -35(-40)\\
$E_4 $ & 3.609(3.586) & -9.489(-9.539) & -29.60(-29.72) & -68(-69) & -80(-70) \\
\hline
\hline 
$^1H_5 $ & 28.61(28.57) & -1.724(-1.727) & -7.925(-7.92) & -32(-32.5) & -60(-60)\\
\hline 
$^3H_5 $ & 6.128(6.082) & -6.41(-6.45) & -24.9(-25.1) & -87(-86) & ---(---)\\
\hline 
$^3G_5 $ & -0.0090(-0.010) & 0.48(0.481) & 1.9
(1.874) & 6(6) & ---(---) \\
$^3I_5 $ & 10.68(10.66) & 0.0108(0.0107) & $0.145(0.143) \,10^{-5}$ & 1.43(1.4) & 6.4(6)\\
$E_5 $ & -31.34(-31.29) & 1.554(1.553) & 6.99
(7.018) & 28(28) & 50(50)
\end{tabular}
\end{ruledtabular}
\end{table*}

\begin{table*}
\caption{\label{tab:table1bis} Contributions ${\bf
\hat M}_0$, ${\bf \hat M}_1$, ${\bf \hat M}_2$, ${\bf \hat M}_3$ and
${\bf \hat M}_4$ for all partial waves in np scattering for the NijmII
and Reid93 (in brackets) potentials at a {\it center of mass momentum
of $m_{\pi}/2$} (equivalent to a laboratory energy about 10 MeV).
From them one can compute the M-matrix at $k_{cm} = m_{\pi}/2$ using 
$
{\bf \hat M}_{l,l'} = ({\bf \hat M_0})_{l,l'} + ({\bf \hat M_1 })_{l,l'} +
({\bf \hat M_2 })_{l,l'} + ({\bf \hat M_3 })_{l,l'} + ({\bf \hat M_4 })_{l,l'}
+ \dots $ 
where $ {\bf M} = {\bf D} \left({\bf S} - 1 \right)
\left({\bf S} +1 \right)^{-1} {\rm i} k {\bf D} $, ${\bf S}$ is the
unitary S-matrix in coupled channel space and ${\bf D}={\rm diag} (
k^{l_1} , \dots, k^{l_N} ) $ is the coupled channel centrifugal
factor. ${\bf \hat M_0} = -{\bf a}^{-1}$, ${\bf \hat
M_1} = \frac{1}{2}{{\bf r_0}}\,(m_{\pi}/2)^2$, ${\bf
\hat M_2} = {{\bf v_2}}\,(m_{\pi}/2)^4$, ${\bf \hat
M_3} = {{\bf v_3}}\,(m_{\pi}/2)^6$, ${\bf \hat
M_4} = {{\bf v_4}}\,(m_{\pi}/2)^8$, and so on.  We have
taken $m_{\pi} = 138.0 {\rm MeV}$.  A long dash (---) stands for cases
where numerical accuracy was outraged and no value could be reliably
deduced.}
\begin{ruledtabular}
\begin{tabular}{|c|c|c|c|c|c|}
\hline 
Wave  & $ M_0 $ & $ M_1 $ & $ M_2 $ & $ M_3 $ & $ M_4 $  \\ 
 & & & & &  \\ 
 & NijmII \,( Reid93 )  & NijmII \,( Reid93 )   
 & NijmII \,( Reid93 )  & NijmII \,( Reid93 ) & NijmII\, ( Reid93 )  \\ 
\hline 
\hline $^1S_0 $ 
& 0.0422(0.0421) & 0.1633(0.1683) 
& -0.0071(-0.0074) & 0.0072(0.0067) 
& -0.0044(-0.0041) \\
\hline $^3P_0 $ 
& 0.4052(0.4050) & 0.2353(0.2366) & 0.0164(0.0144) & 0.0070(0.0068) 
& -0.0017(-0.0017) \\
\hline
\hline 
$^1P_1 $ 
& -0.3575(-0.3655) & -0.3912(-0.4039) & -0.0236(-0.0274) 
& 0.0007(0.0019) & 0.0017(0.0019) \\
\hline 
$^3P_1 $ 
& -0.6539(-0.6538) & -0.5245(-0.5230) 
& -0.0003(-0.0002) & -0.0016(-0.0017)
&  0.0001(0.00005) \\
\hline 
$^3S_1 $ 
& -0.1999(-0.1999) & 0.1121(0.1121) 
& -0.0020(-0.0021) & 0.0027(0.0026) 
& -0.0017(-0.0017) \\
$^3D_1 $ 
& -0.1666(-0.1680) & -0.2154(-0.2180) & -0.0553(-0.0569) 
&  0.0021(0.0019)  & -0.0006(-0.0006) \\
$E_1 $ 
& 0.05062(0.05097) & 0.02467(0.02524) 
& -0.00410(-0.00395) & 0.00266(0.00261) 
& -0.00164(-0.0016) \\
\hline 
\hline 
$^1D_2 $ 
& 0.720(0.726) & 0.909(0.919) & 0.245(0.250) & -0.024(-0.024) 
& 0.008(0.008) \\
\hline 
$^3D_2 $ 
& 0.1350(0.1349) & 0.1747(0.1743) & 0.0358(0.0354) 
& -0.0018(-0.0018) & 0.0005(0.0005) \\
\hline 
$^3P_2 $ 
& -0.4222(-0.4257) & -0.5056(-0.511) & -0.1033(-0.1065) 
& -0.011(-0.012) & -0.005(-0.01) \\
$^3F_2 $ 
& -0.1230(-0.127) & -0.3448(-0.356) & -0.3433(-0.356) 
& -0.1450(-0.152) & -0.026(-0.029) 
\\
$E_2 $ 
& -0.6960(-0.7022) & -0.9601(-0.9712) & -0.3764(-0.3845) 
& -0.043(-0.045) & -0.015(-0.02) \\
\hline
\hline 
$^1F_3 $ 
& -0.1193(-0.1196) & -0.2399(-0.2406) & -0.1478(-0.1486) 
& -0.0281(-0.0285) & -0.001(-0.001) \\
\hline 
$^3F_3 $ 
& -0.3699(-0.3723) & -0.6072(-0.6110) & -0.3073(-0.3099) 
& -0.036(-0.036) & -0.004(-0.005) \\
\hline 
$^3D_3 $ 
& 0.05153(0.05151) & 0.0837(0.08347) & 0.0308(0.0304) 
& 0.004(0.004) & ---(---) \\
$^3G_3 $ 
& $-1.530(-1.870) \, 10^{-3}$ 
& $-2.021(-3.233) \, 10^{-3}$ 
& $-0.248(-1.885) \, 10^{-3}$ 
& $-0.21(-1.21) \, 10^{-3}$ 
& $-0.7(-0.9) \, 10^{-3}$ 
\\
$E_3 $ 
& 0.1024(0.1023) & 0.1990(0.1986) & 0.1144(0.1139) 
& 0.0174(0.0170) & ---(---) \\
\hline 
\hline 
$^1G_4 $ 
& 0.3087(0.3115) & 0.6587(0.6616) & 0.514(0.516) & 0.15(0.15) 
& 0.02(---) \\
\hline 
$^3G_4 $ 
& 0.0522(0.0522) & 0.1257(0.1259) & 0.1018(0.1020) 
& 0.0304(0.0305) & 0.003(0.003) \\
\hline 
$^3F_4 $ 
& -0.0961(-0.0965) & -0.185(-0.186) & -0.12(-0.12) & -0.02(---) 
& ---(---) \\
$^3H_4 $ 
& -0.00080(-0.00082) & -0.01077(-0.01088) 
& -0.0227(-0.0228) & -0.020(-0.020) 
& -0.01(-0.007) \\
$E_4 $ 
& -0.2774(-0.2791) & -0.5801(-0.5831) & -0.443(-0.445) 
& -0.125(-0.125) & -0.02(-0.02) \\
\hline
\hline 
$^1H_5 $
& -0.0350(-0.0350) & -0.1054(-0.1055) & -0.118(-0.119) & -0.059(-0.059) 
& -0.015(-0.02) \\
\hline 
$^3H_5 $ 
& -0.1632(-0.1644) & -0.392(-0.394) & -0.373(-0.38) & -0.16(-0.15) 
& ---(---) \\
\hline 
$^3G_5 $ 
& 0.0109(0.0109) & 0.0294(0.0294) & 0.028(0.028) & 0.01(0.012) 
& ---(---) \\
$^3I_5 $ 
& $-0.0092(-0.0105)\,10^{-3}$ 
& $0.661(0.65)\,10^{-3}$ 
& $2.2(2.2)\,10^{-3}$  
& $2.6(2.6)\,10^{-3}$  
& $1.5(1.4)\,10^{-3}$  
\\
$E_5 $ 
& 0.0319(0.0319) & 0.0950(0.0951) & 0.1045(0.1046) & 0.051(0.051) & 0.01(0.011) \\
\end{tabular}
\end{ruledtabular}
\end{table*}

\subsection{Validity of the effective range expansion for partial waves}
\label{sec:ere} 

\begin{figure*}
\begin{center}
\epsfig{figure=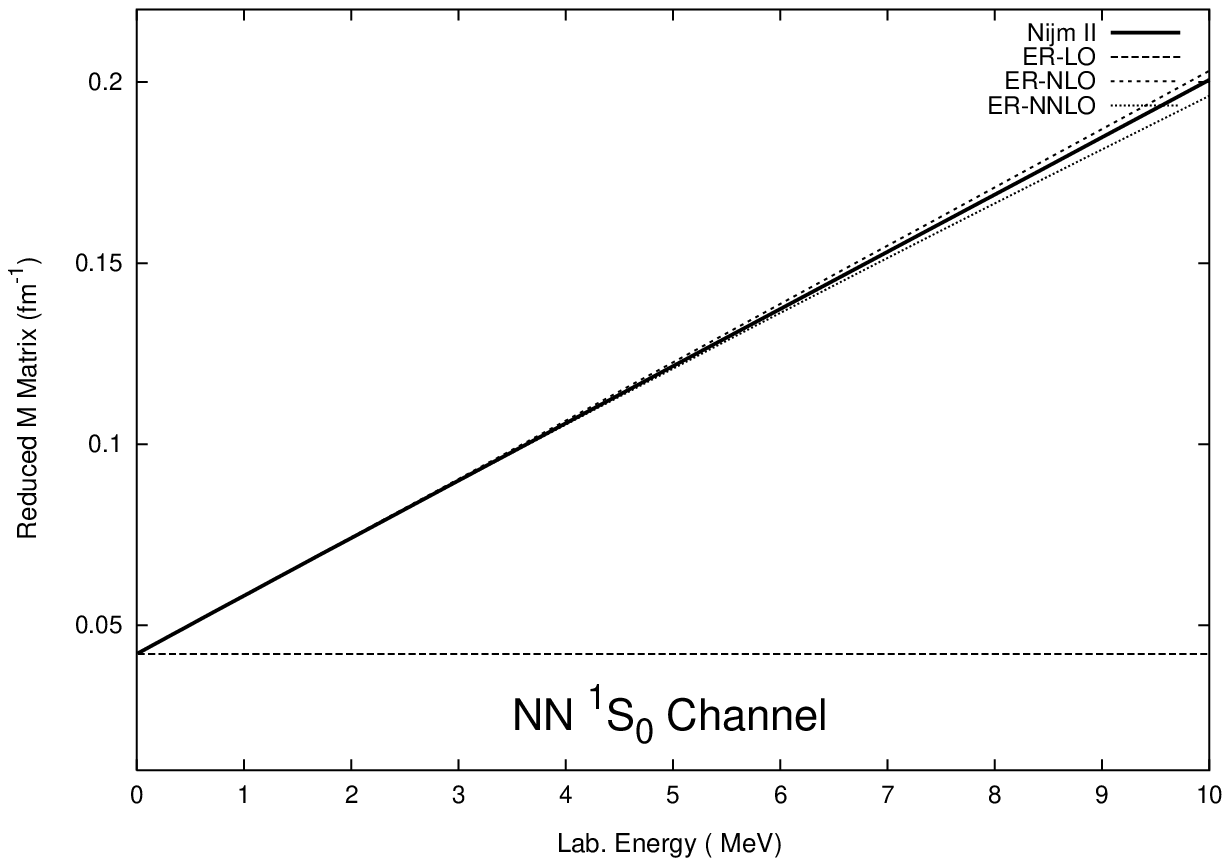,height=8cm,width=8cm}
\epsfig{figure=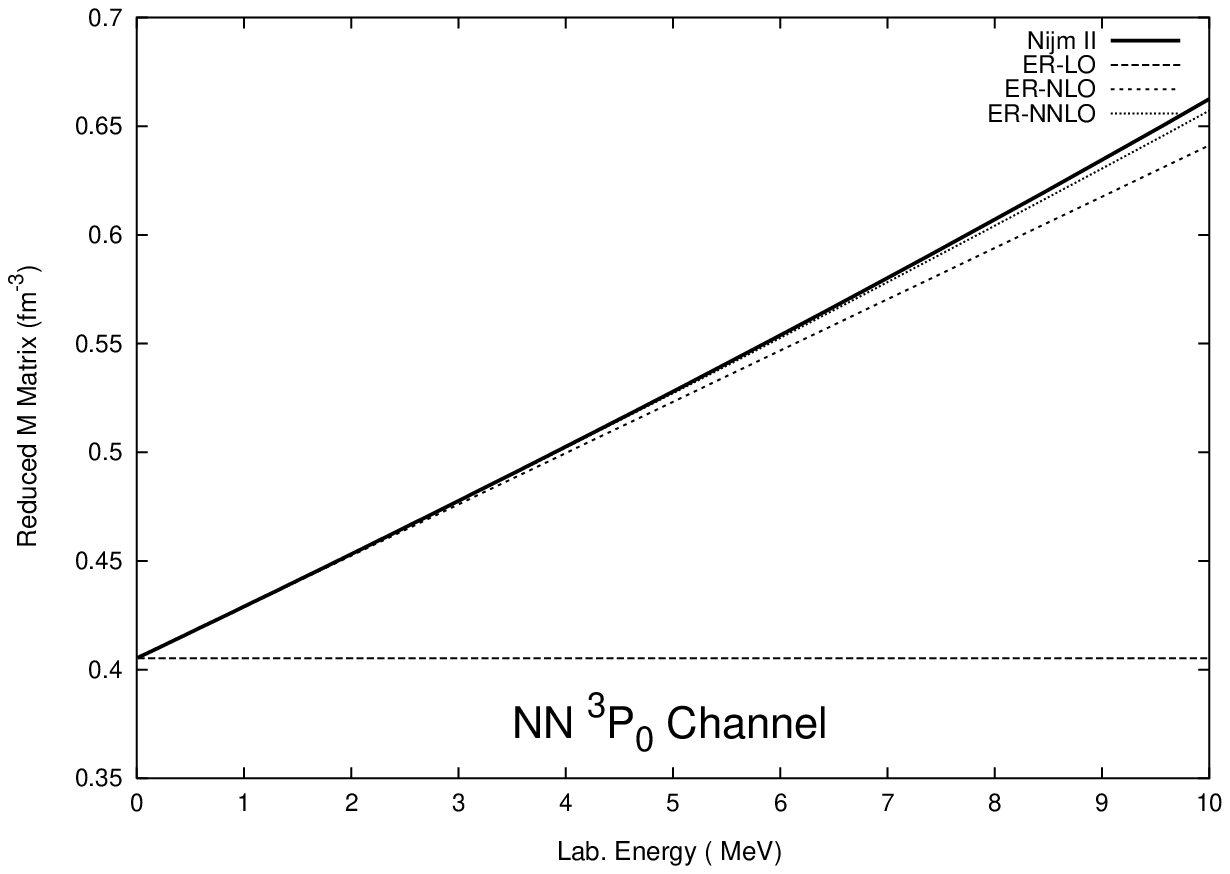,height=8cm,width=8cm}
\end{center}
\caption{np scaled M-matrix on the effective range expansion for
states with total angular momentum $j=0$. We construct the NijmII
scales M-matrix corresponding to the generalized effective range
expansion of Eq.~(\ref{eq:c-ere}) to a given order as a function of
the LAB energy.  LO means including order $k^0$ terms, NLO including
$k^2$ terms and NNLO including $k^4$ terms with the low energy
threshold parameters obtained from solving the evolution equations for
the threshold parameters, Eq.~(\ref{eq:var}) with the NijmII potential
. Data are obtained from the partial waves phase shift analysis from
Ref.~\cite{Stoks:1993tb}.}
\label{fig:M-matrix_j=0}
\end{figure*}

\begin{figure*}
\begin{center}
\epsfig{figure=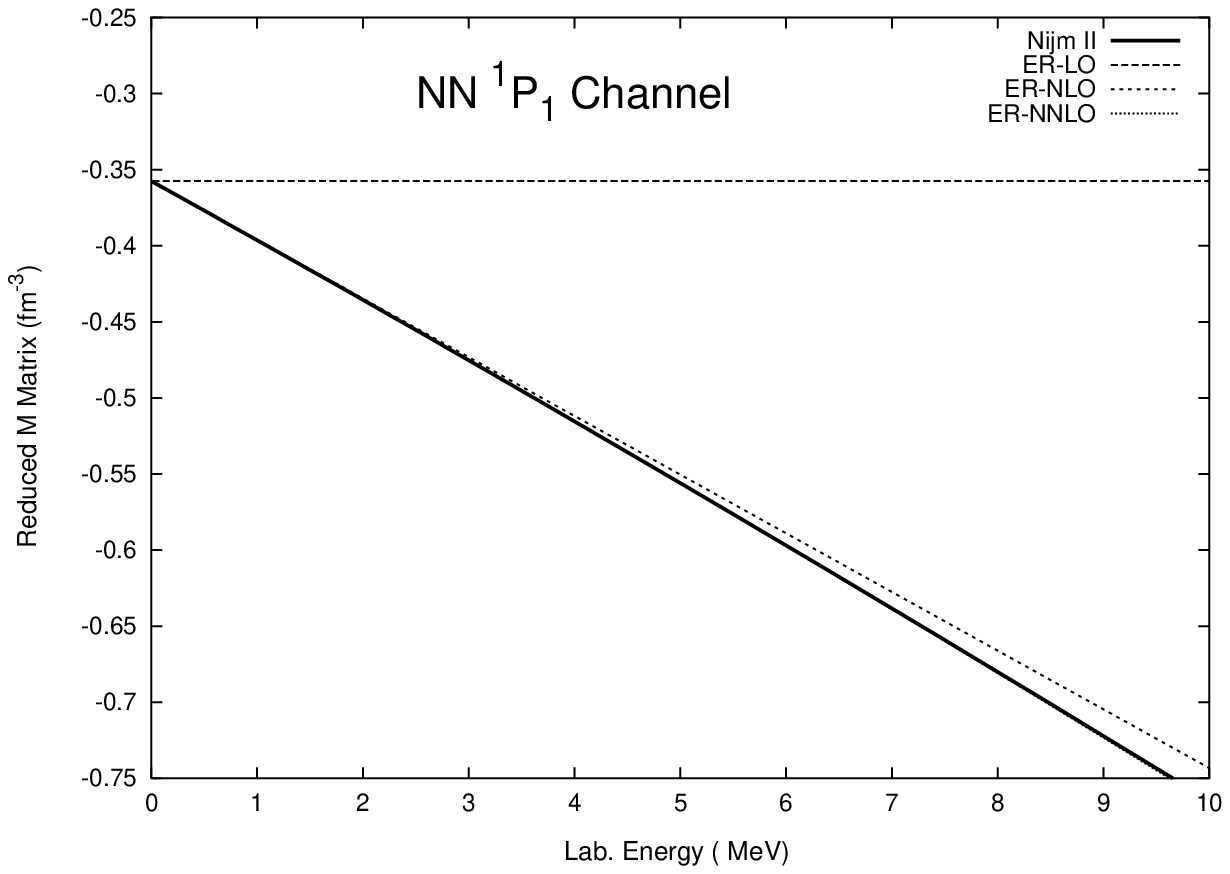,height=8cm,width=8cm} 
\epsfig{figure=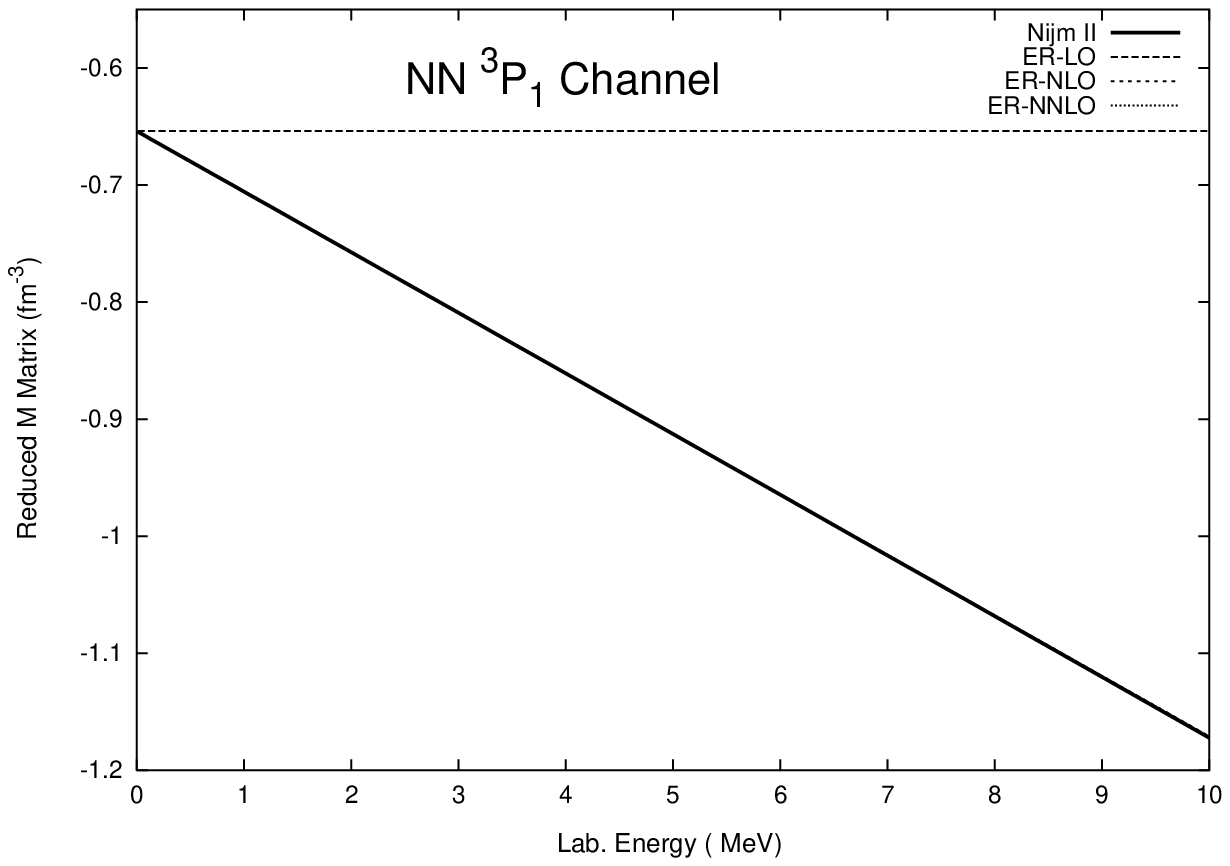,height=8cm,width=8cm} 
\end{center}
\begin{center}
\epsfig{figure=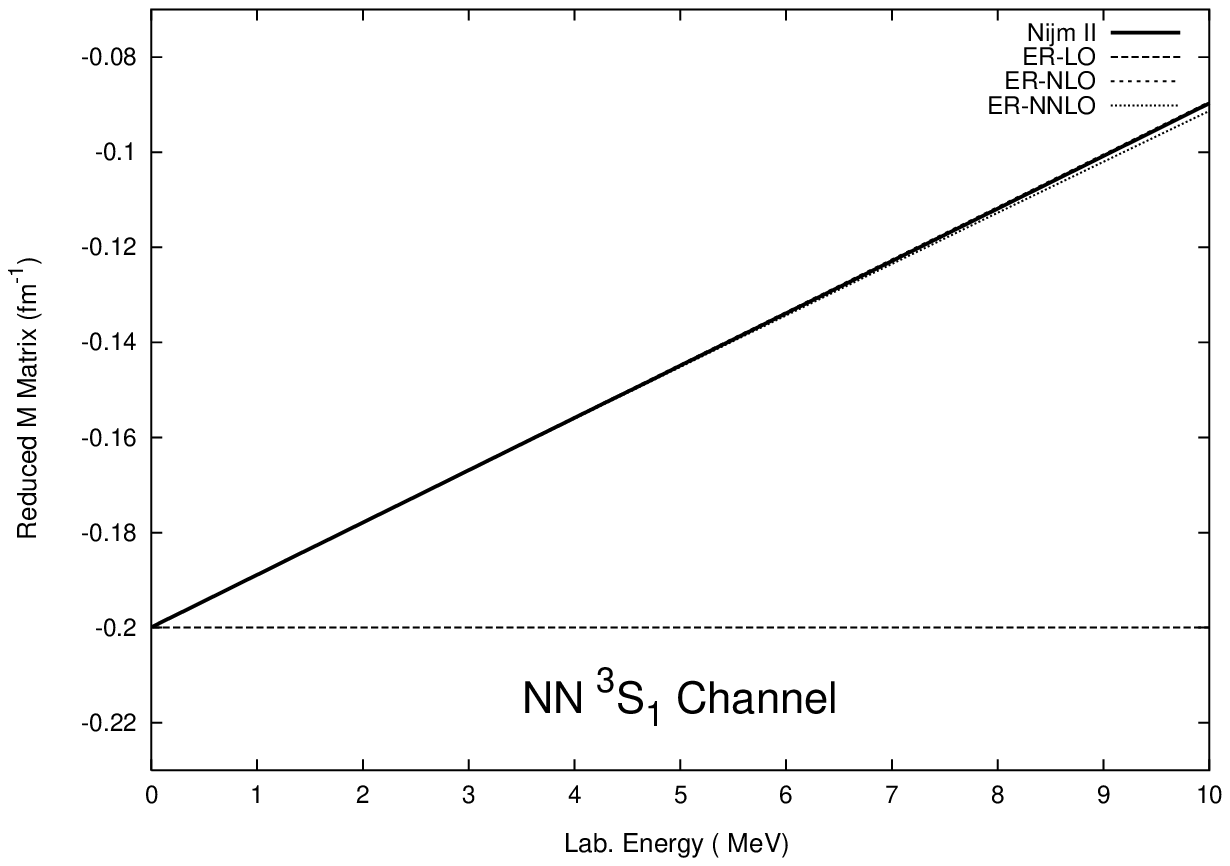,height=8cm,width=8cm} 
\epsfig{figure=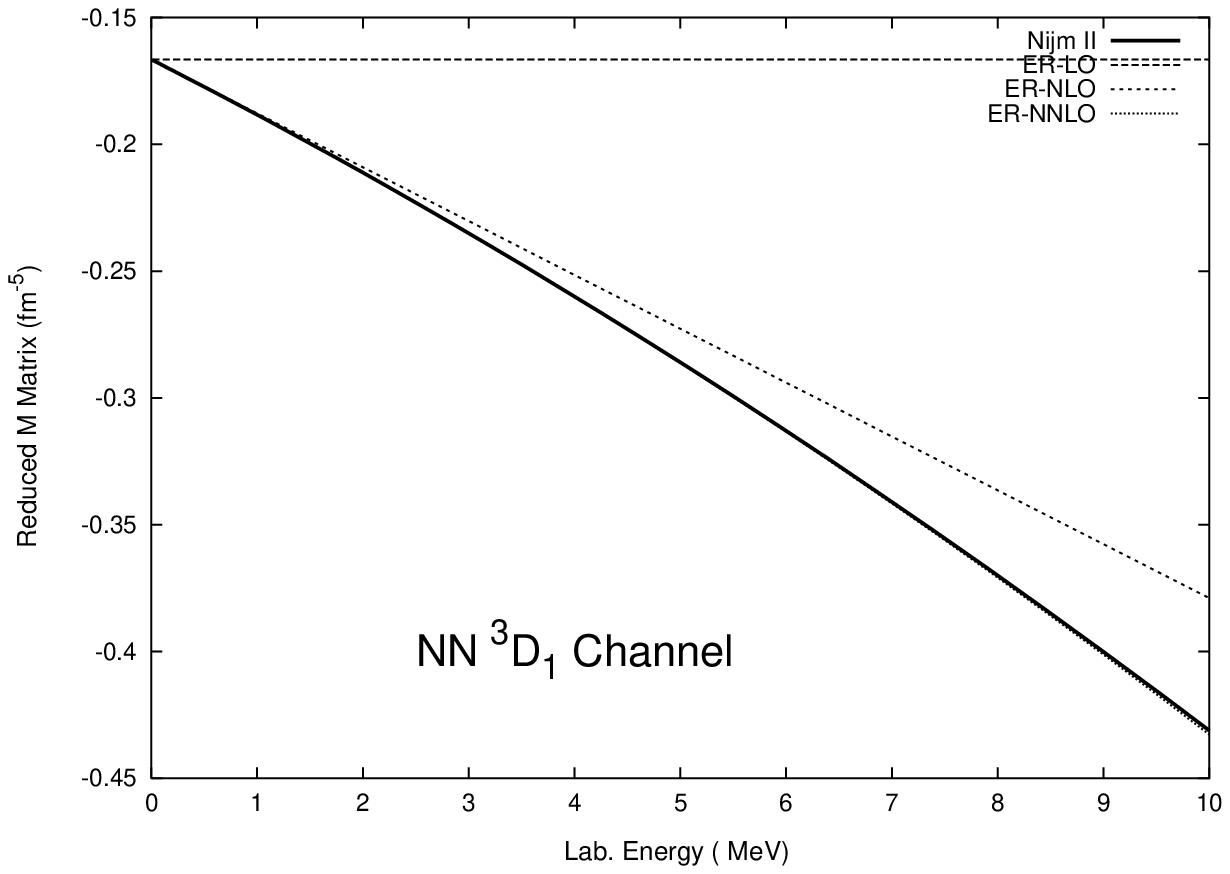,height=8cm,width=8cm} 
\epsfig{figure=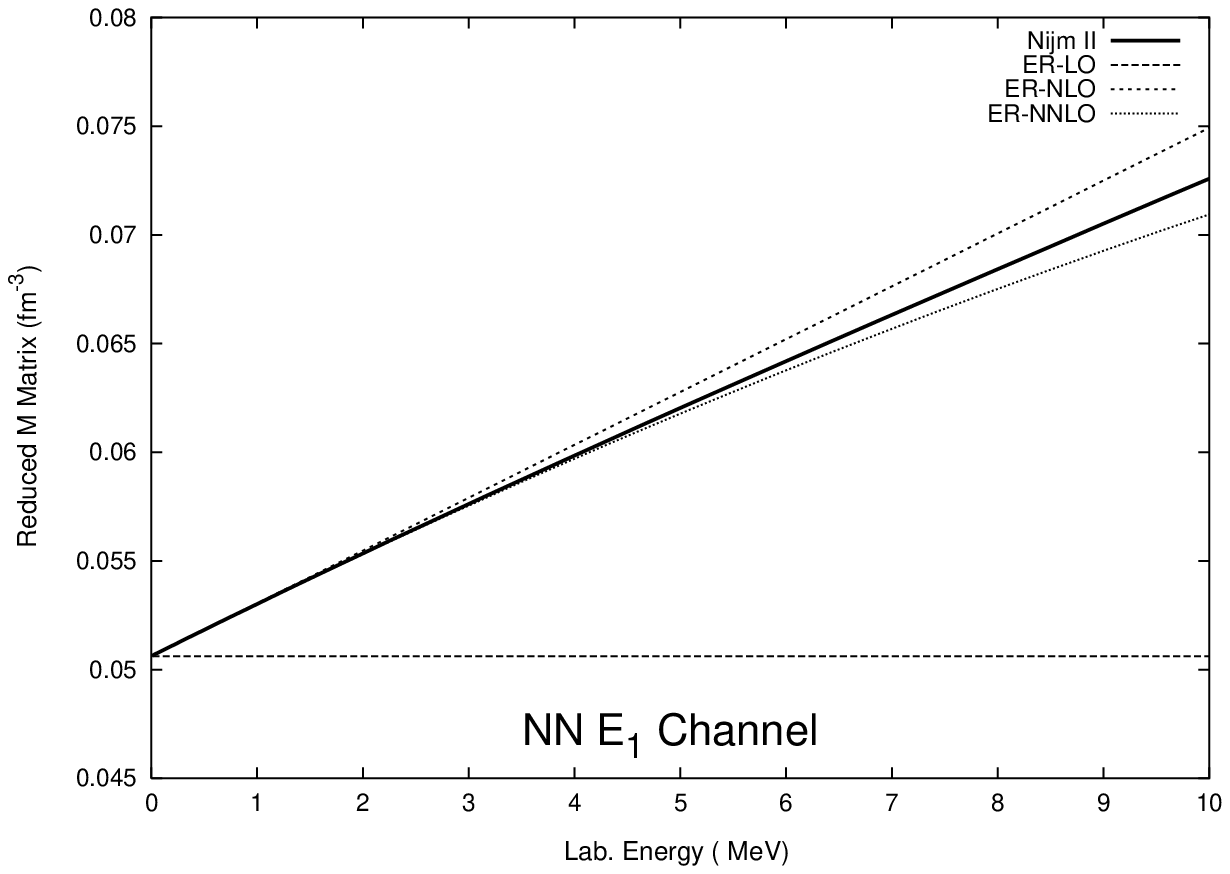,height=8cm,width=8cm} 
\end{center}
\caption{np scaled M-matrix based on the effective range expansion for total
angular momentum $j=1$. For notation see
Fig.~\ref{fig:M-matrix_j=0}. }
\label{fig:M-matrix_j=1}
\end{figure*}

\begin{figure*}
\begin{center}
\epsfig{figure=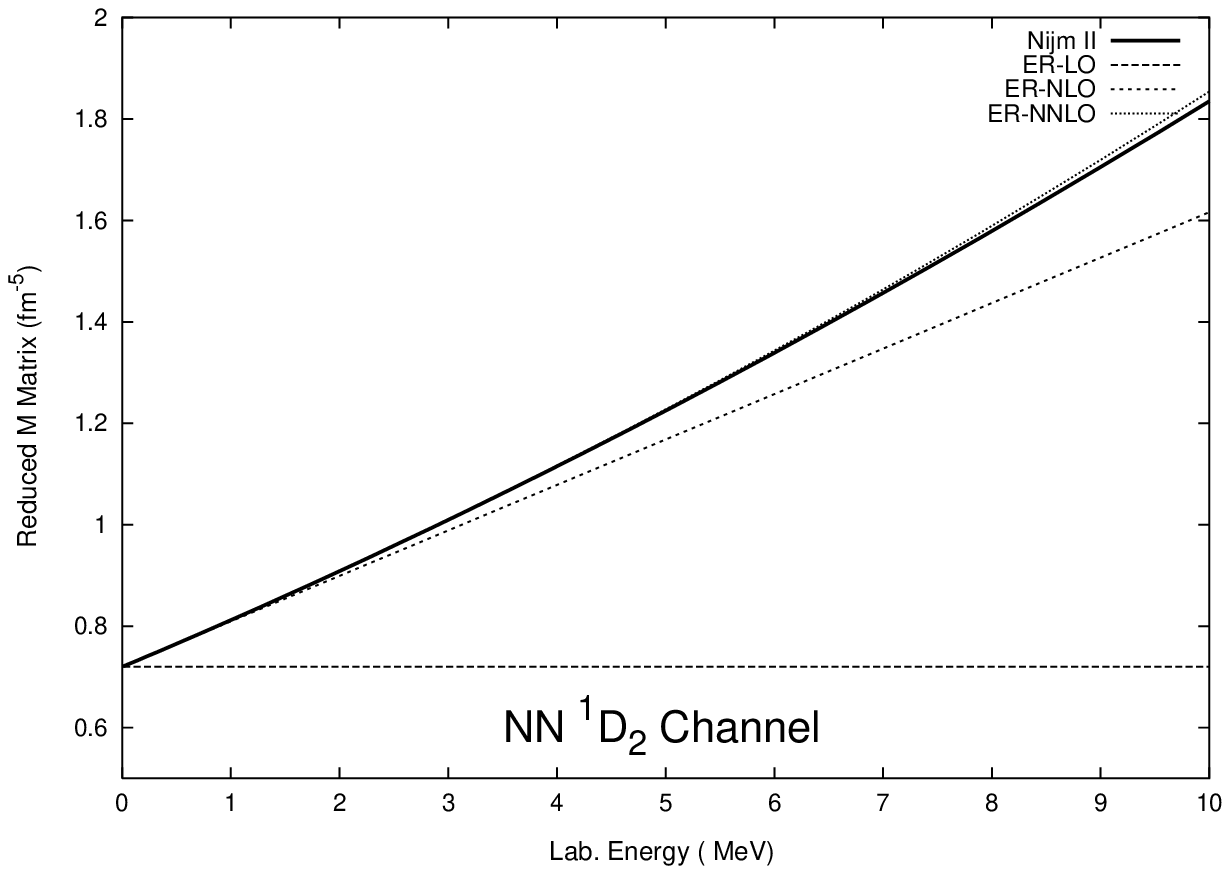,height=8cm,width=8cm} 
\epsfig{figure=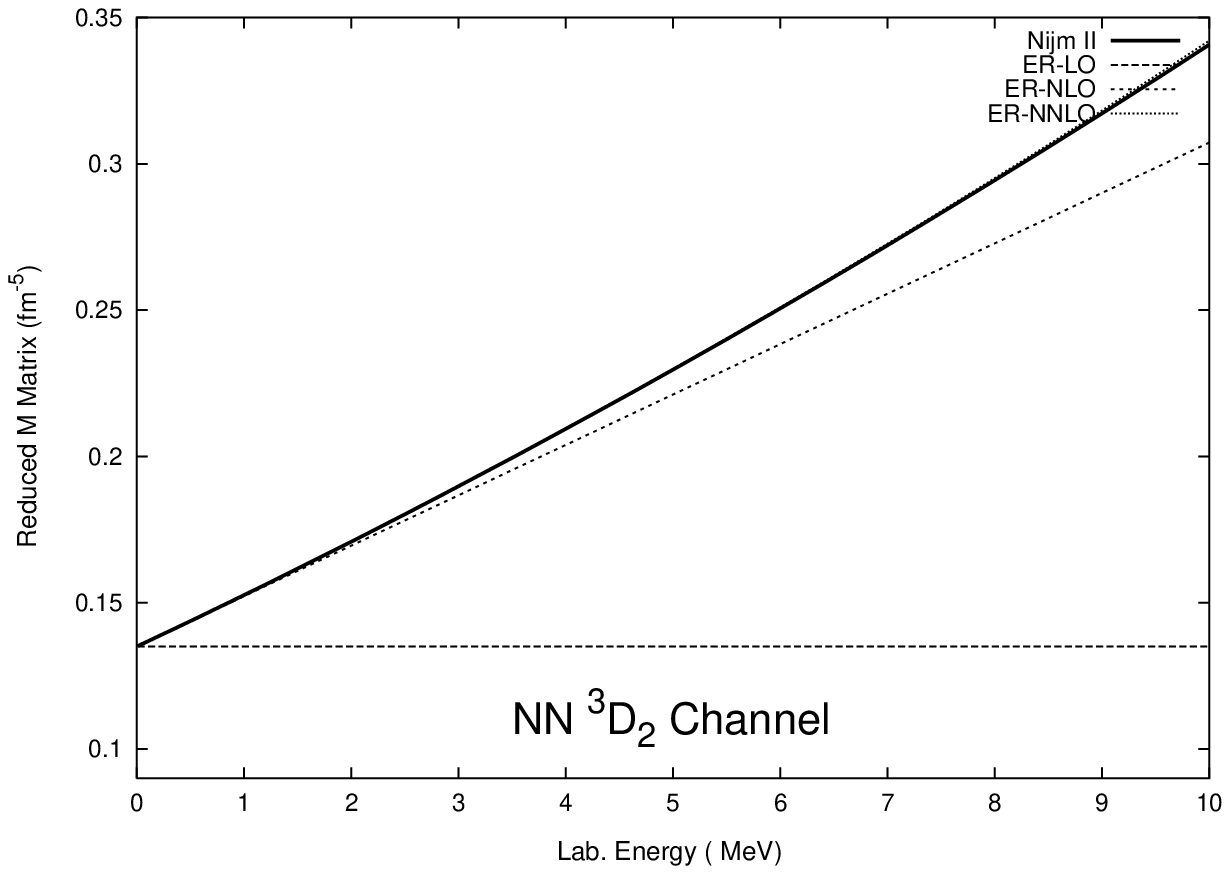,height=8cm,width=8cm} 
\end{center}
\begin{center}
\epsfig{figure=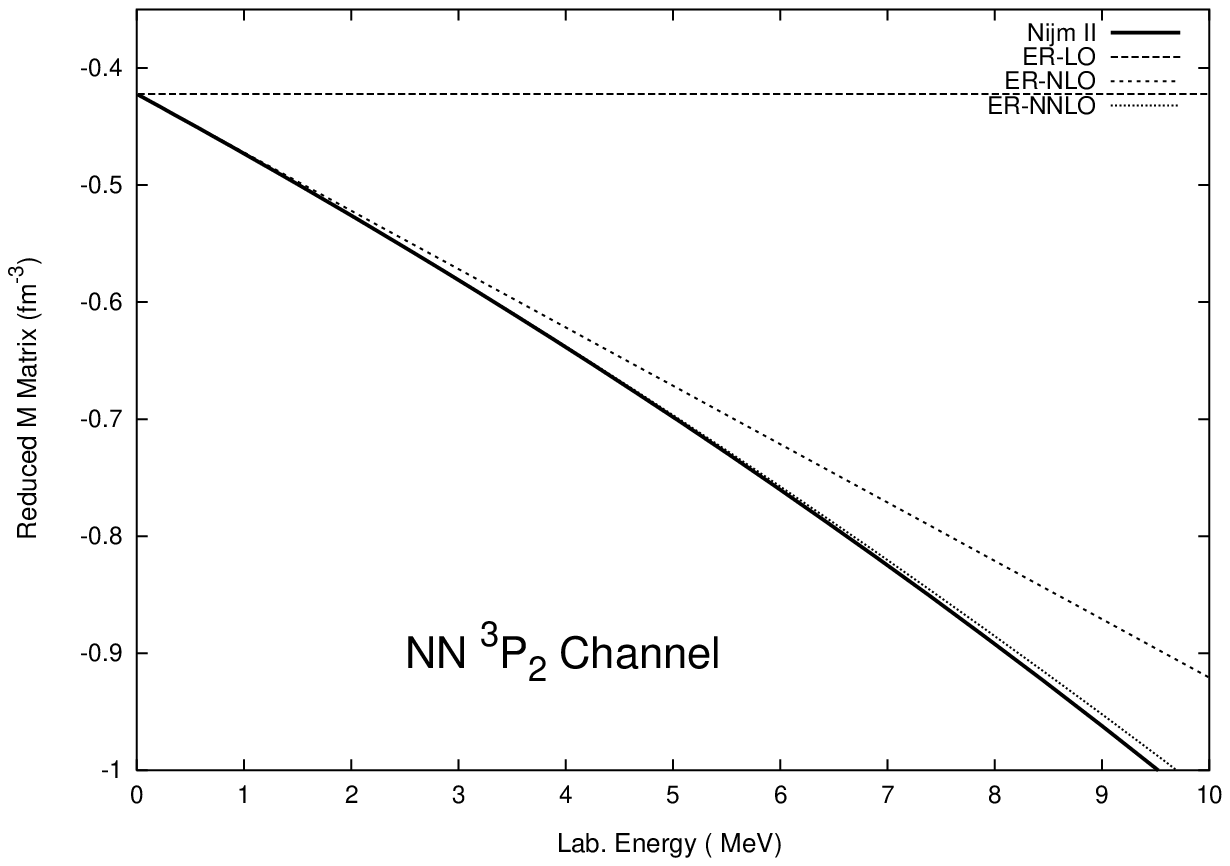,height=8cm,width=8cm} 
\epsfig{figure=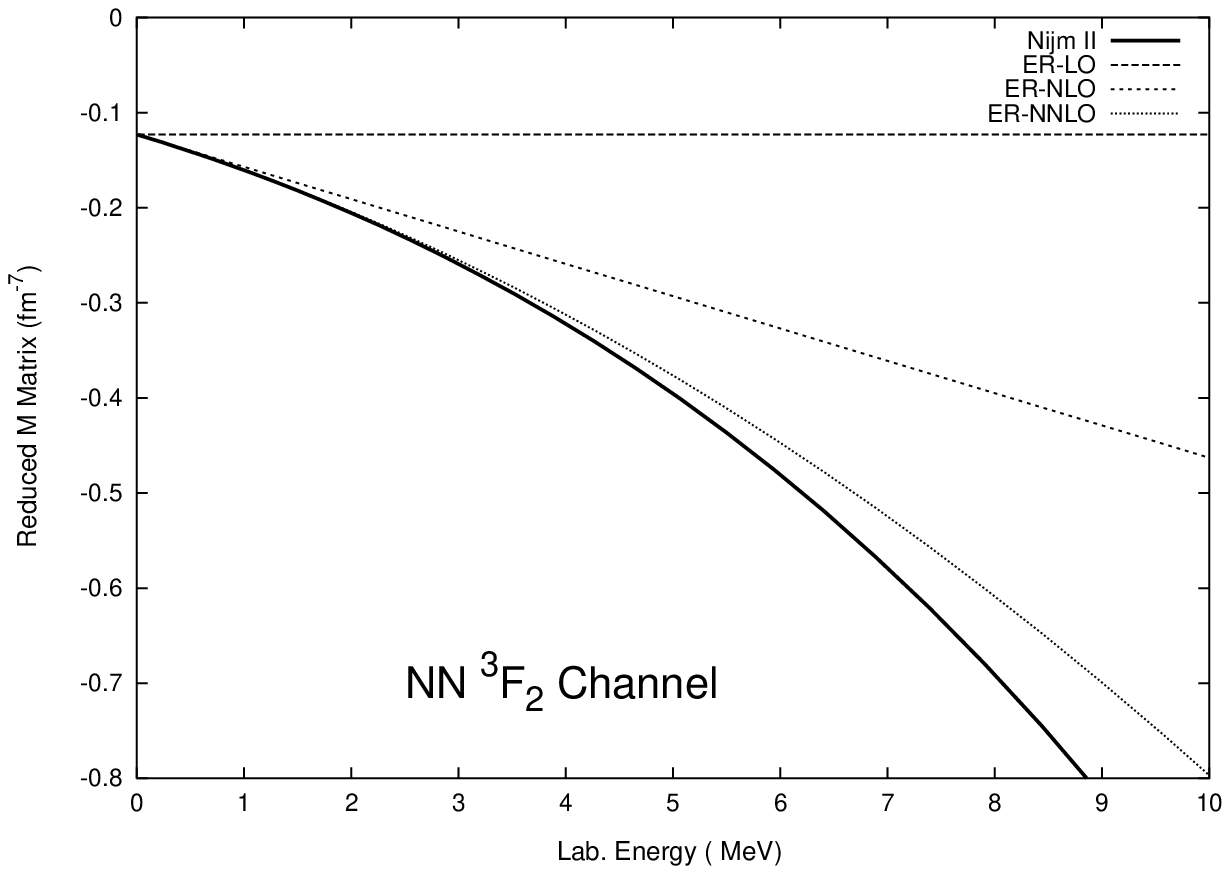,height=8cm,width=8cm} 
\epsfig{figure=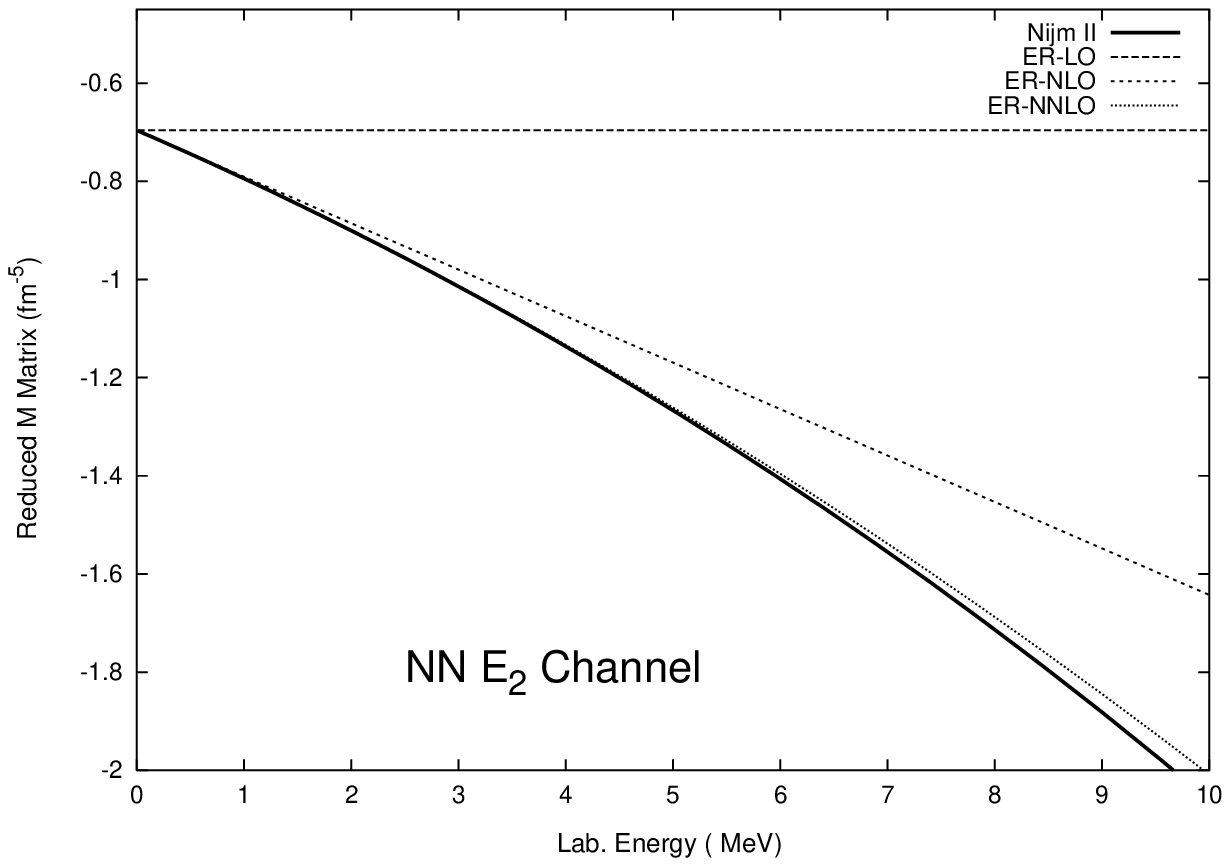,height=8cm,width=8cm} 
\end{center}
\caption{np scaled M-matrix based on the effective range expansion for total
angular momentum $j=2$. For notation see
Fig.~\ref{fig:M-matrix_j=0}. }
\label{fig:M-matrix_j=2}
\end{figure*}

\begin{figure*}
\begin{center}
\epsfig{figure=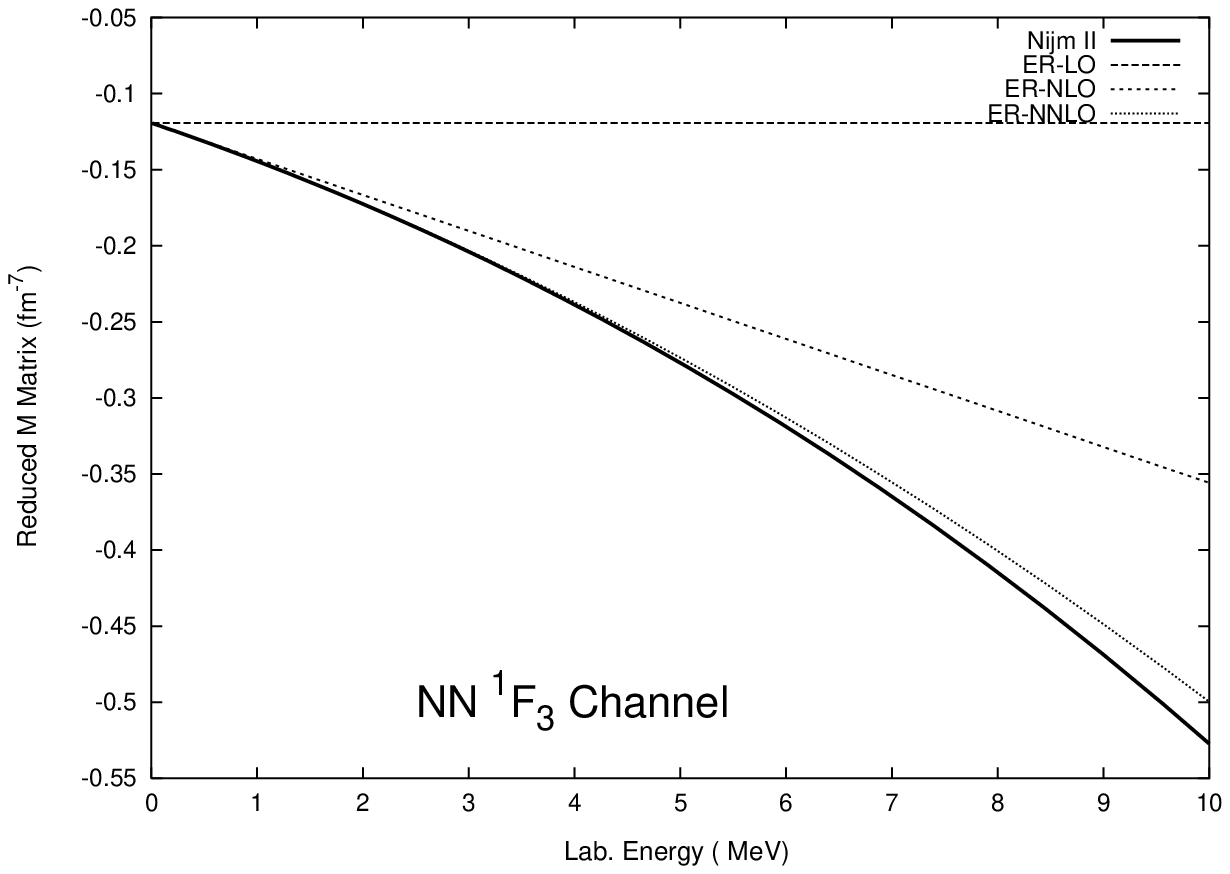,height=8cm,width=8cm} 
\epsfig{figure=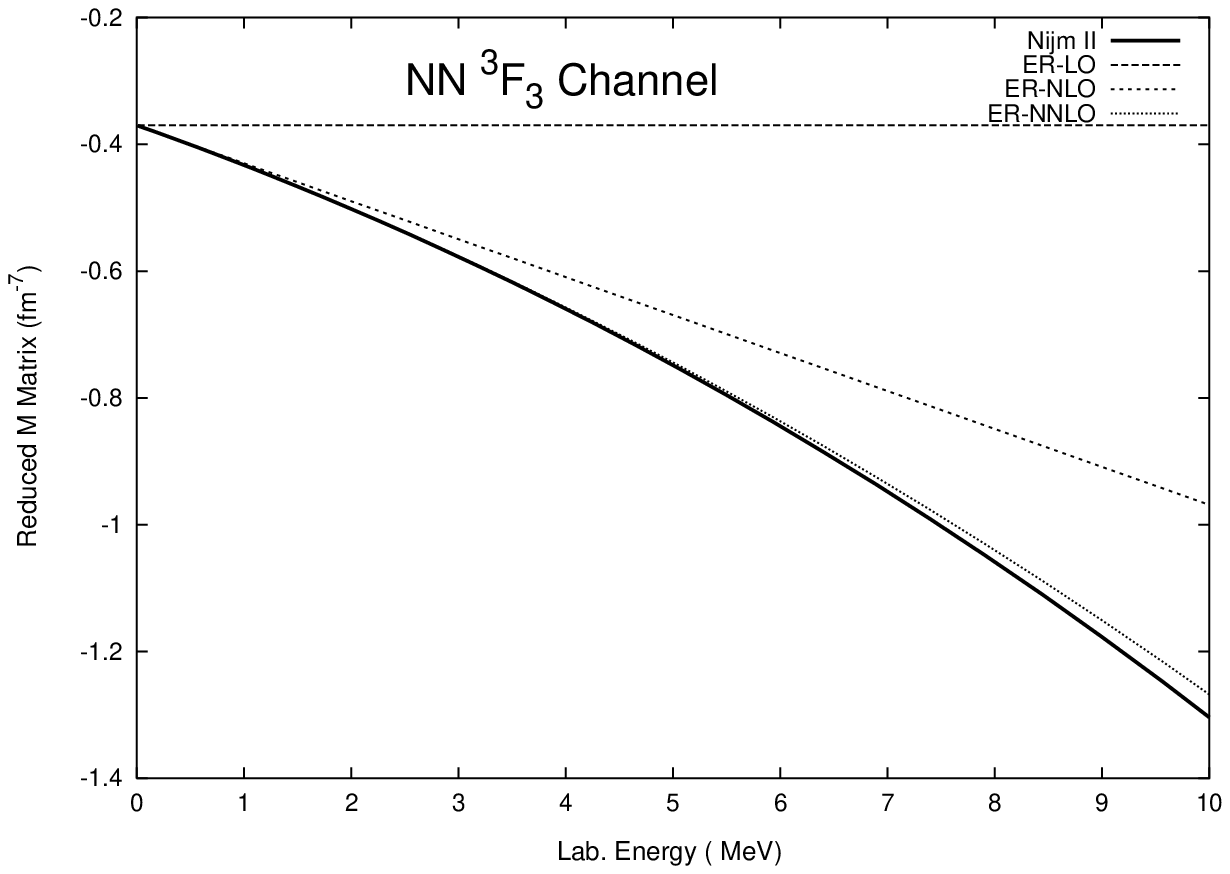,height=8cm,width=8cm} 
\end{center}
\begin{center}
\epsfig{figure=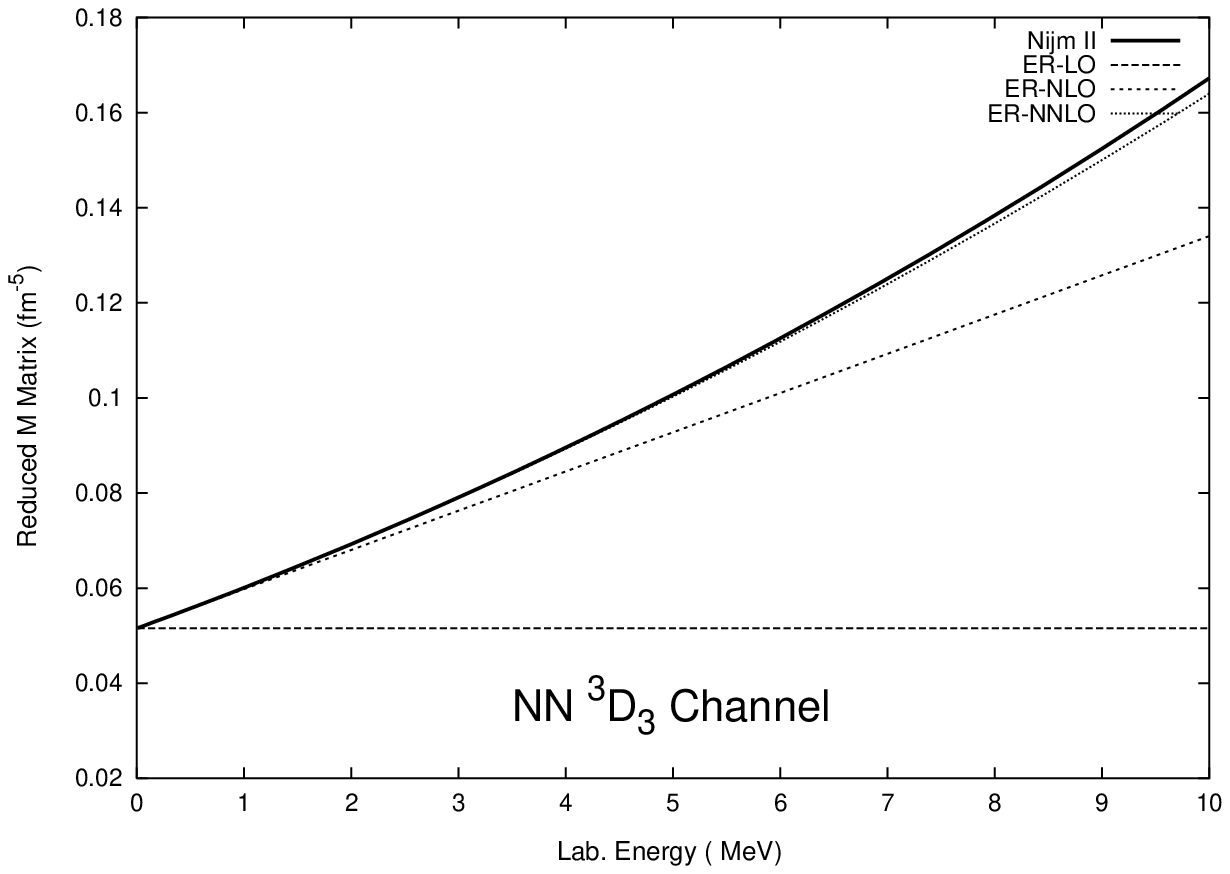,height=8cm,width=8cm} 
\epsfig{figure=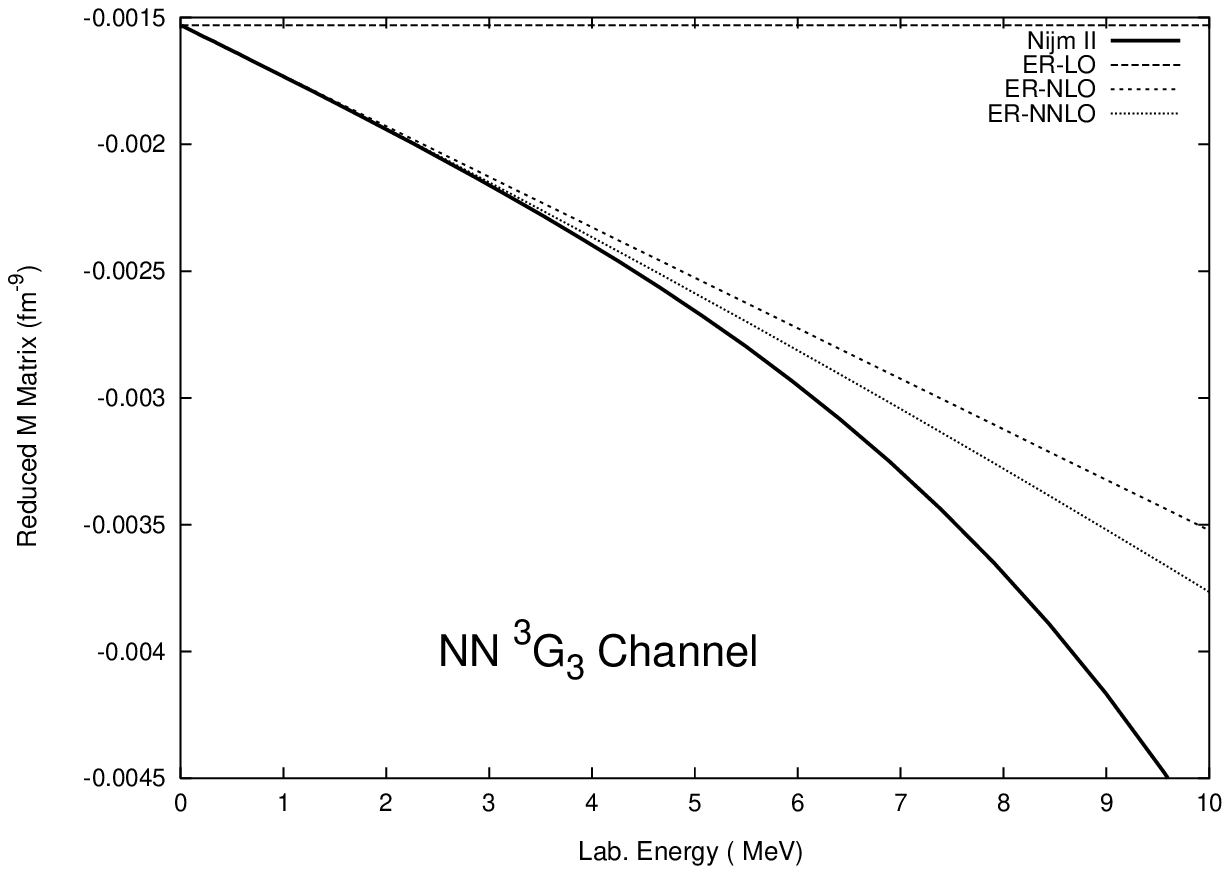,height=8cm,width=8cm} 
\epsfig{figure=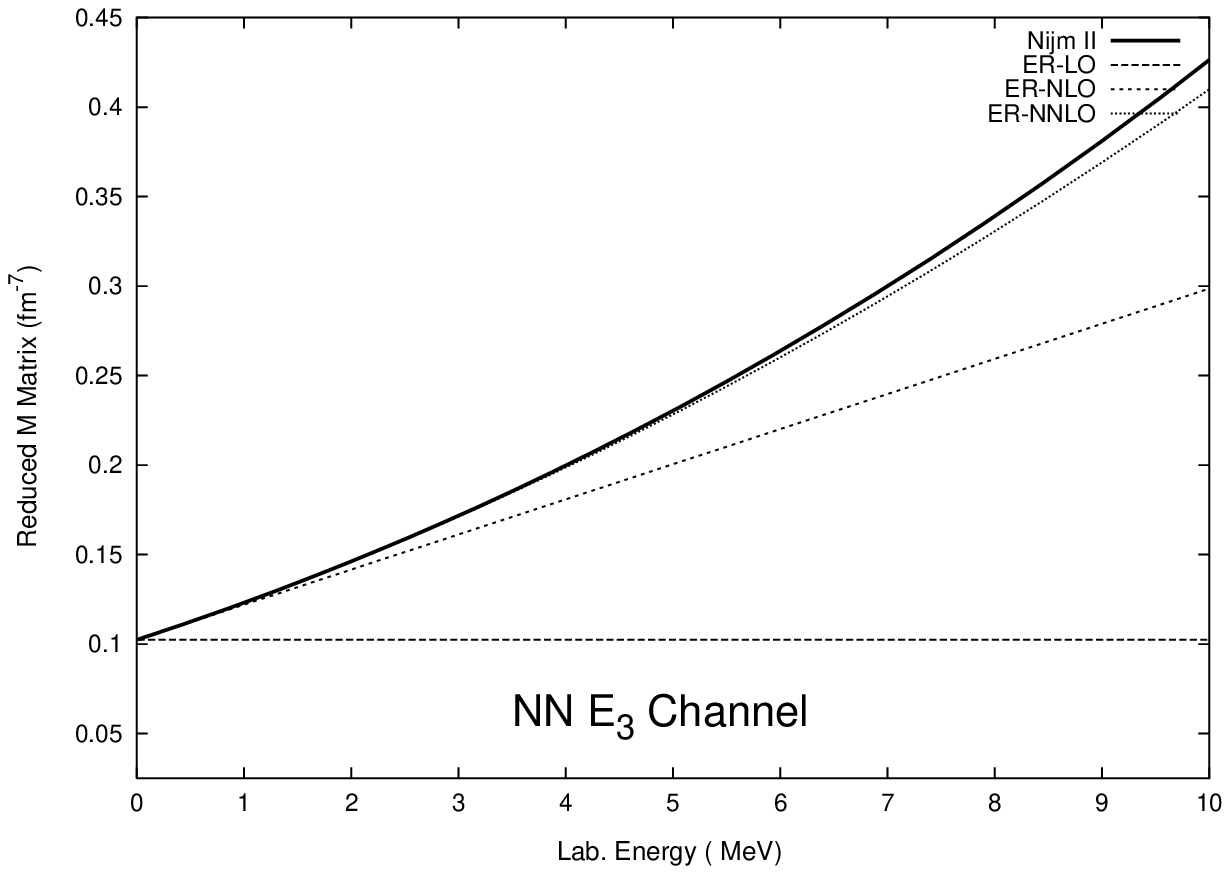,height=8cm,width=8cm} 
\end{center}
\caption{np scaled M-matrix based on the effective range expansion for total
angular momentum $j=3$. For notation see
Fig.~\ref{fig:M-matrix_j=0}. }
\label{fig:M-matrix_j=3}
\end{figure*}

\begin{figure*}
\begin{center}
\epsfig{figure=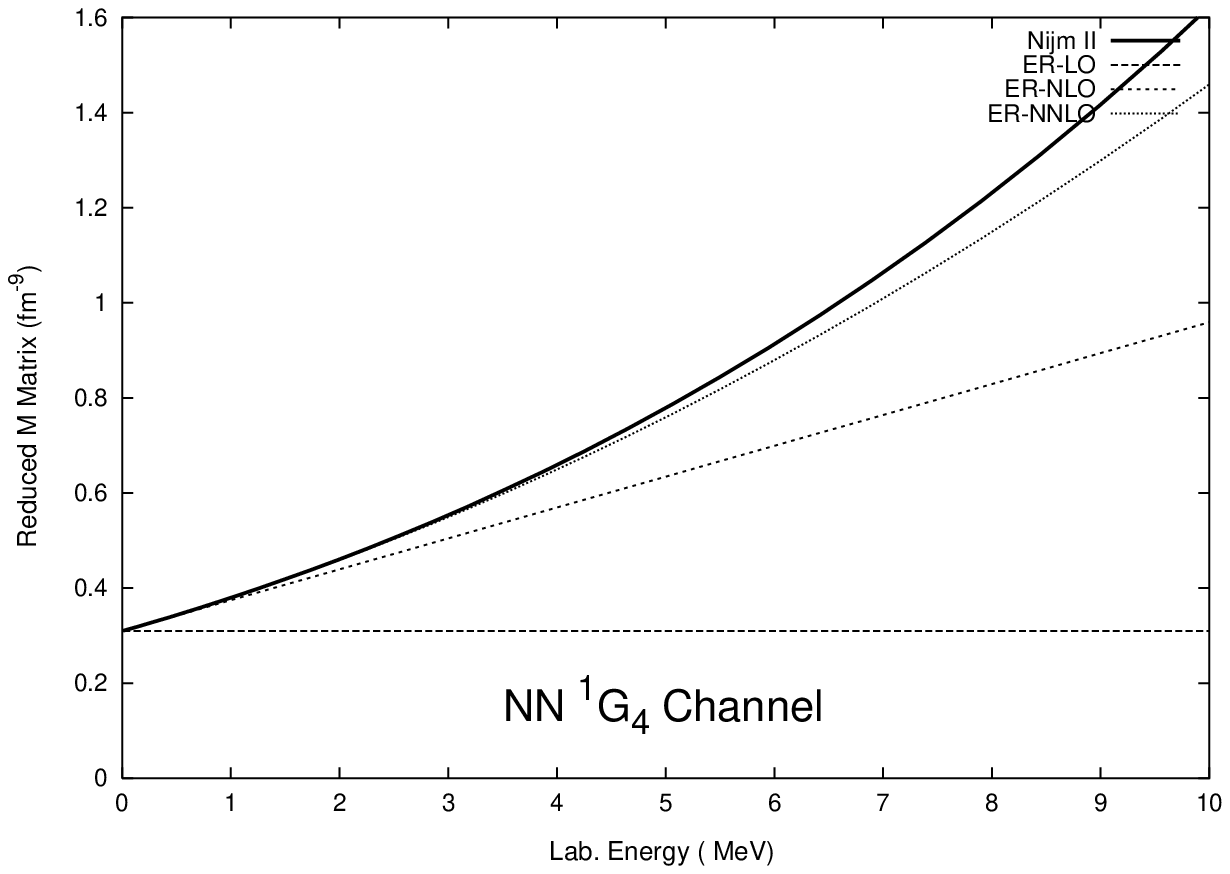,height=8cm,width=8cm} 
\epsfig{figure=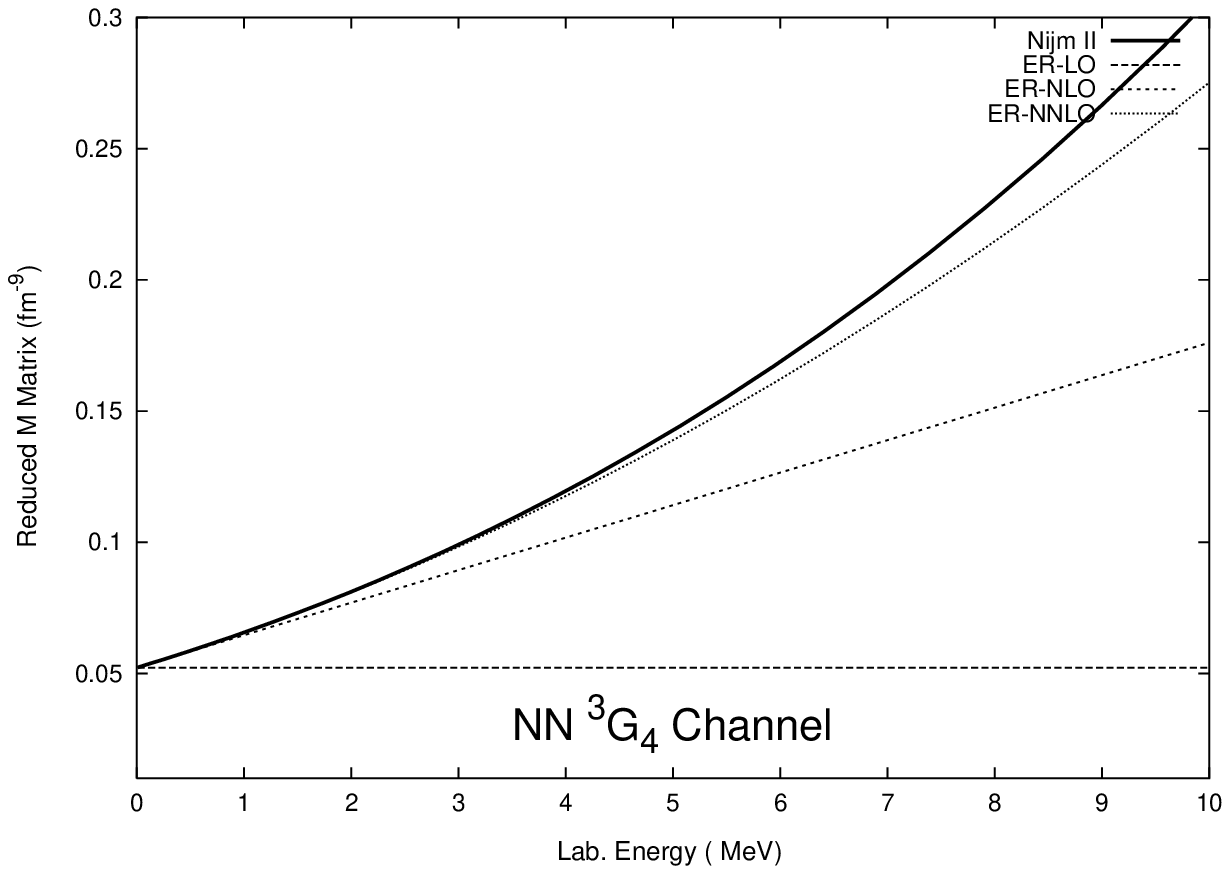,height=8cm,width=8cm} 
\end{center}
\begin{center}
\epsfig{figure=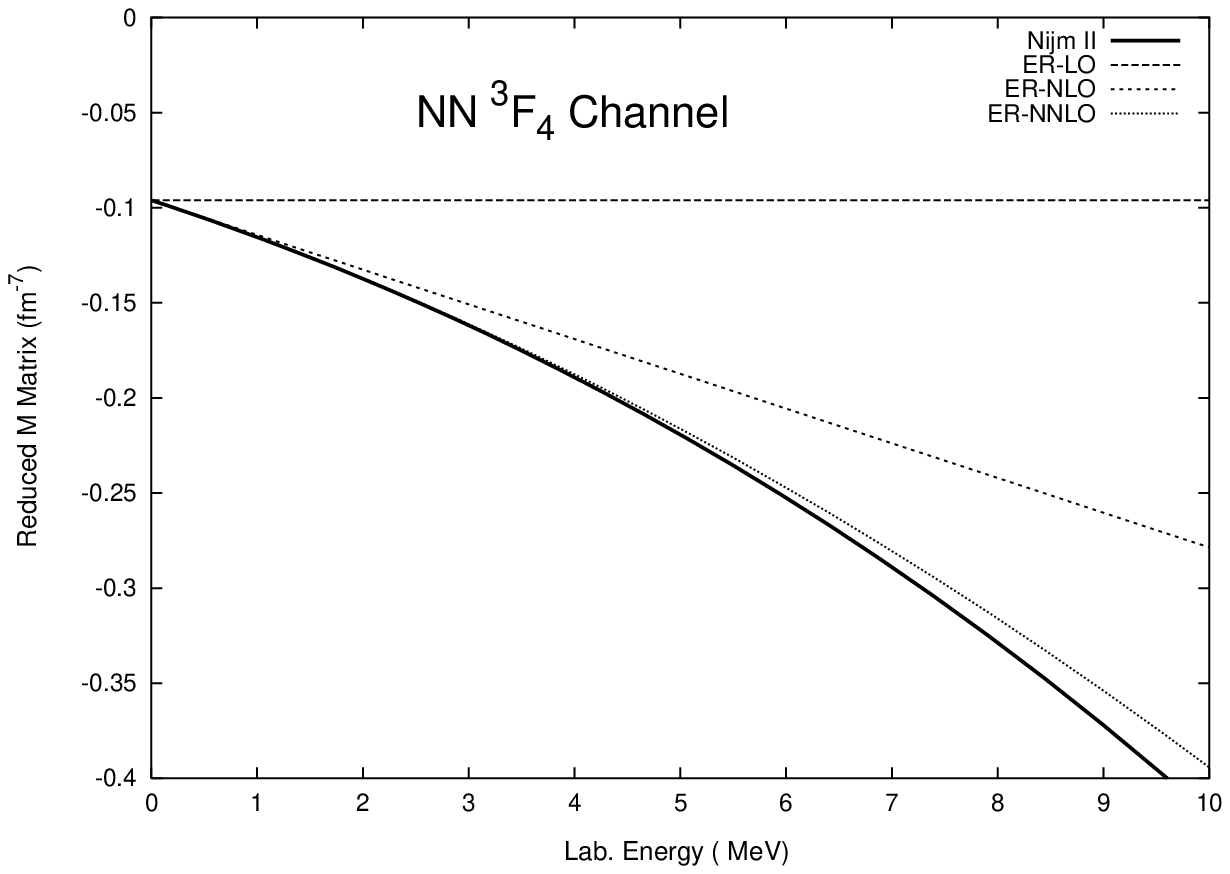,height=8cm,width=8cm} 
\epsfig{figure=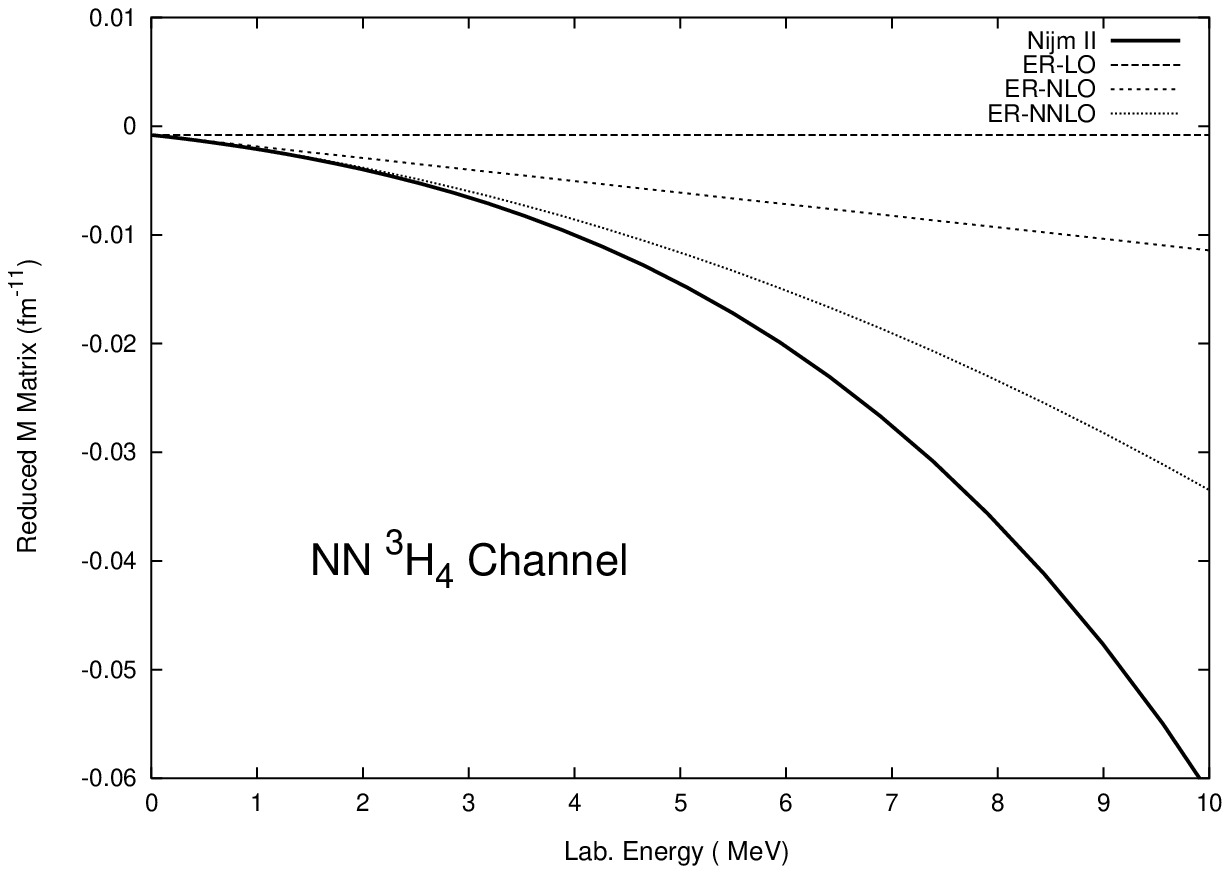,height=8cm,width=8cm} 
\epsfig{figure=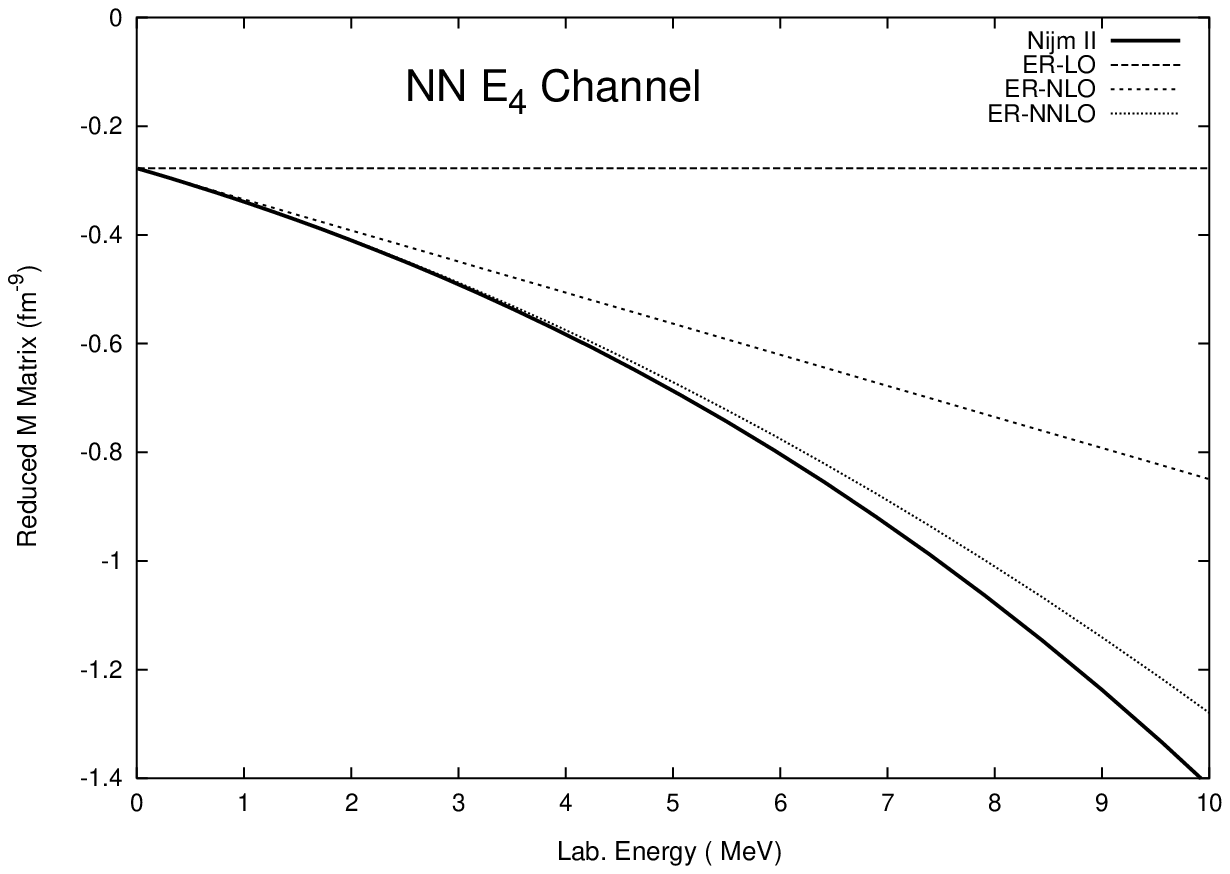,height=8cm,width=8cm} 
\end{center}
\caption{np scaled M-matrix based on the effective range expansion for total
angular momentum $j=4$. For notation see
Fig.~\ref{fig:M-matrix_j=0}. }
\label{fig:M-matrix_j=4}
\end{figure*}

\begin{figure*}
\begin{center}
\epsfig{figure=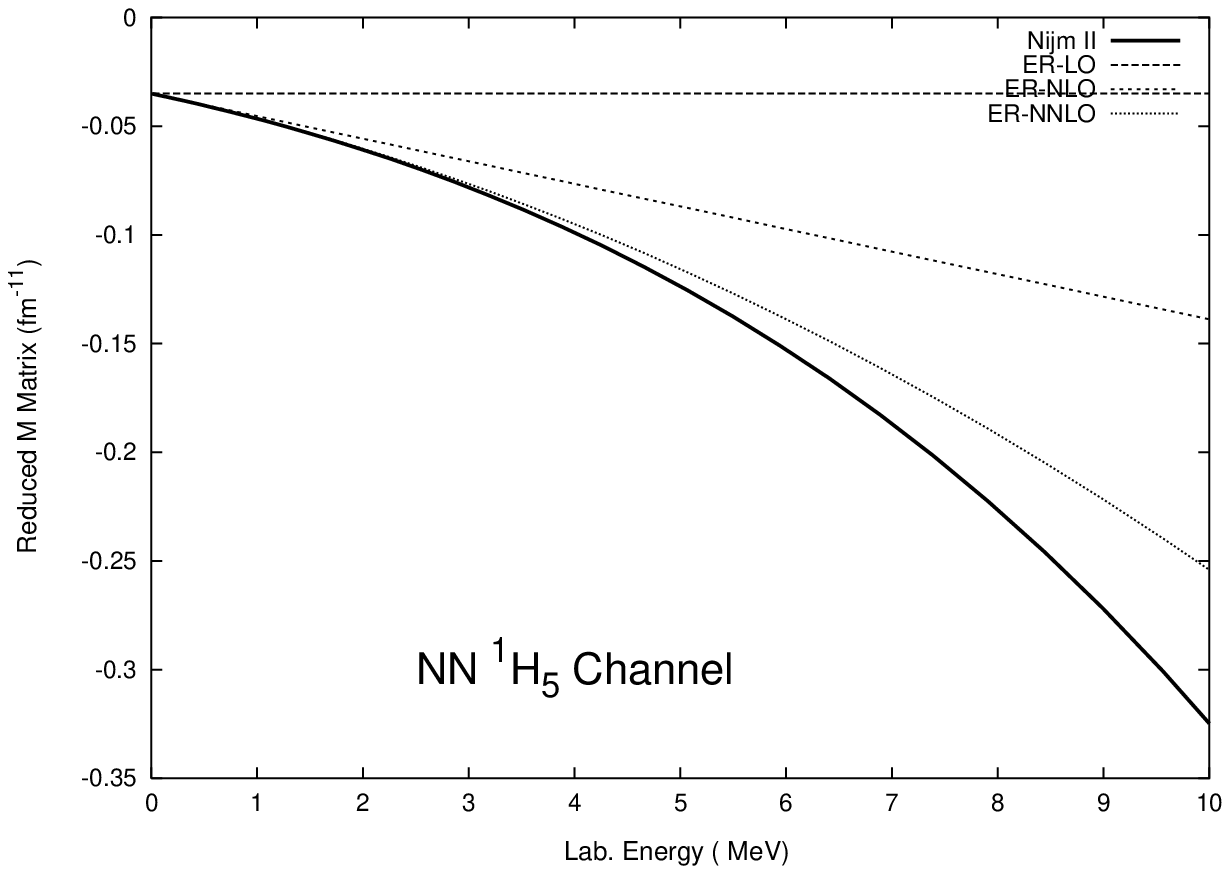,height=8cm,width=8cm} 
\epsfig{figure=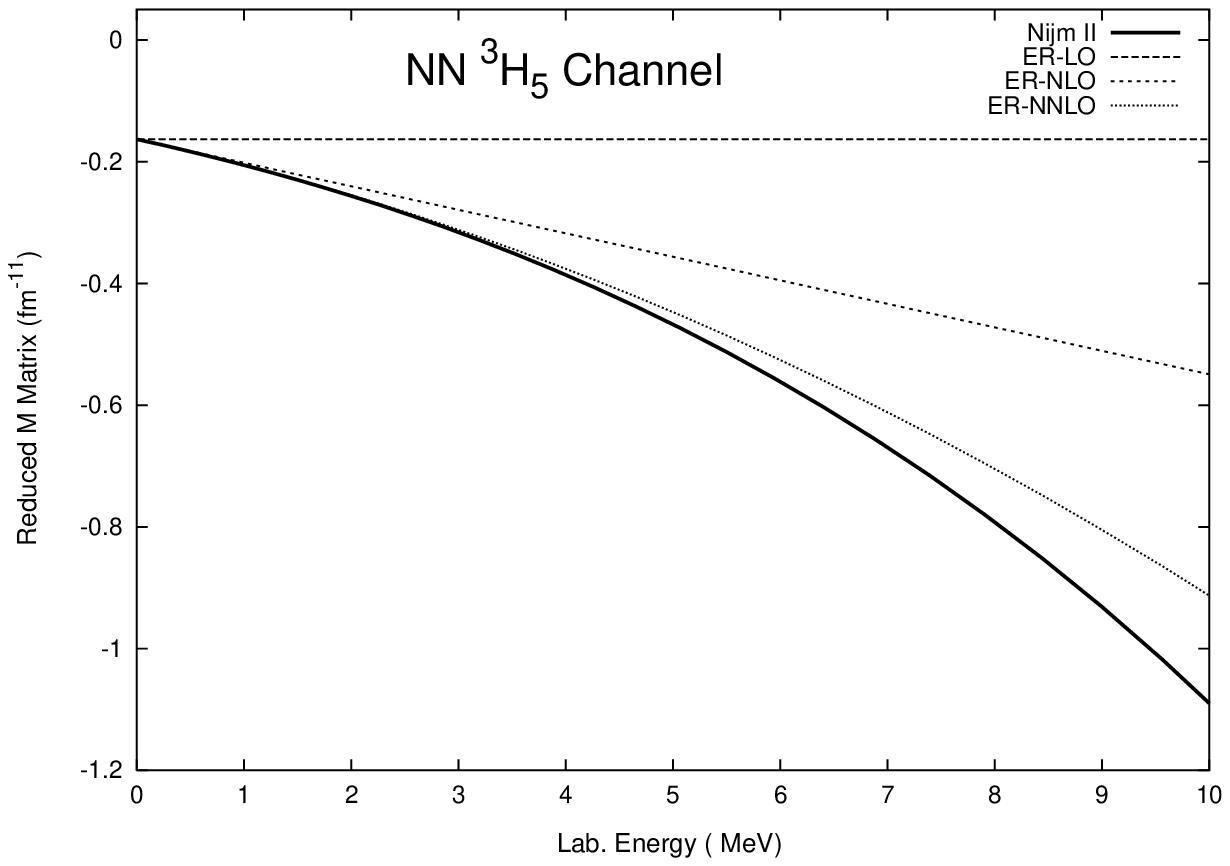,height=8cm,width=8cm} 
\end{center}
\begin{center}
\epsfig{figure=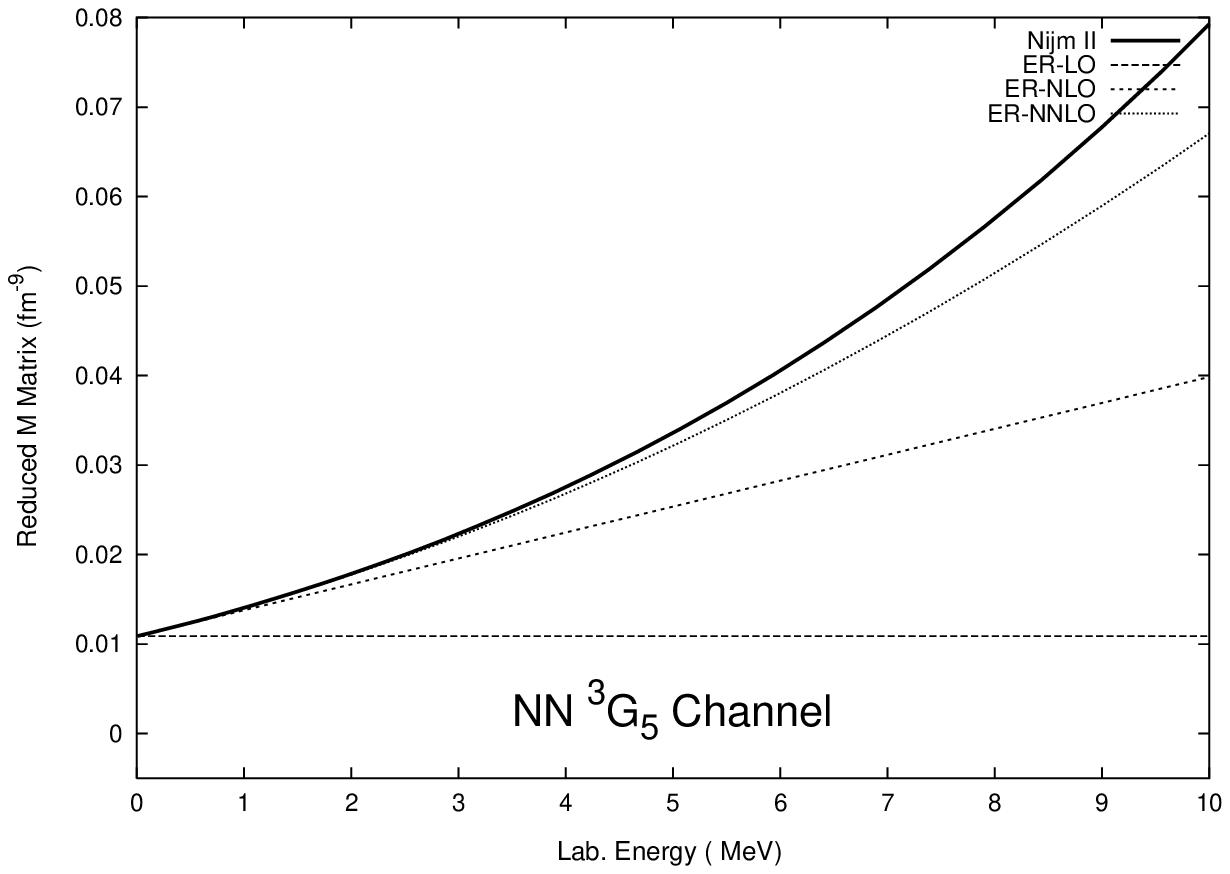,height=8cm,width=8cm} 
\epsfig{figure=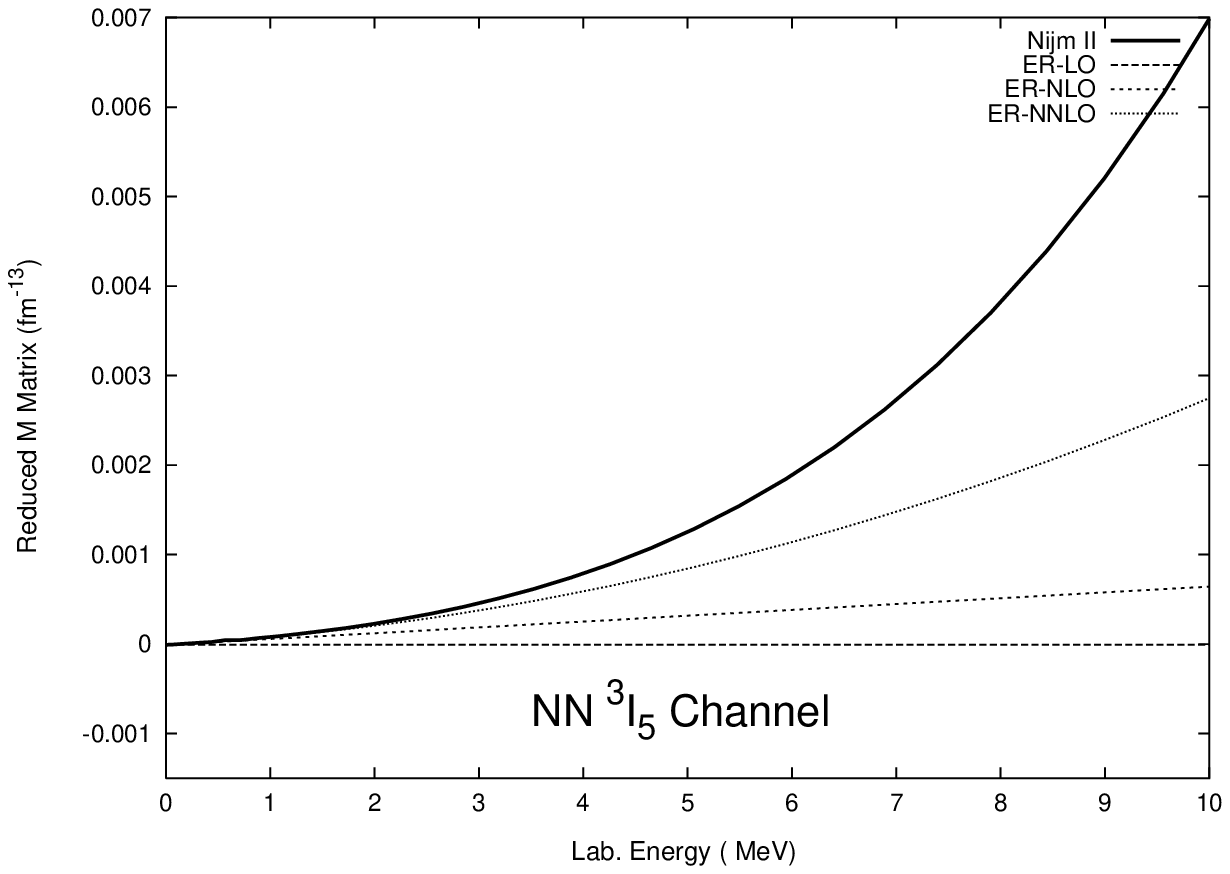,height=8cm,width=8cm} 
\epsfig{figure=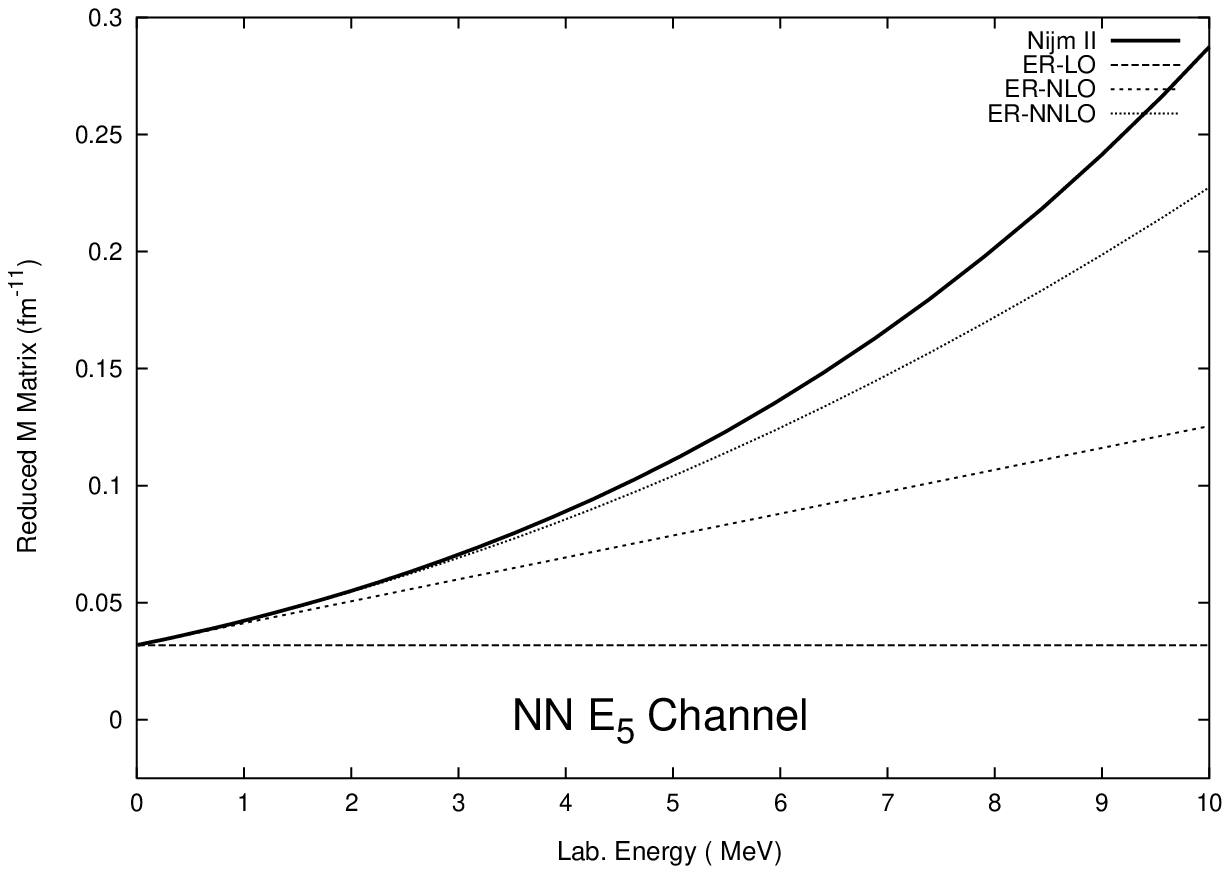,height=8cm,width=8cm} 
\end{center}
\caption{np scaled M-matrix based on the effective range expansion for total
angular momentum $j=5$. For notation see
Fig.~\ref{fig:M-matrix_j=0}. }
\label{fig:M-matrix_j=5}
\end{figure*}

Once the low energy threshold parameters have been determined we can
readily check the validity of the effective range expansion given by
Eq.~(\ref{eq:c-ere}). On theoretical grounds the full expansion around
the origin should be convergent within the analyticity domain, i.e. up
to the next singularity, which for NN interaction happens at about $ k
= \pm {\rm i} m_\pi / 2 $, i.e. $E_{\rm LAB} \sim -10 {\rm MeV} $
where it develops a logarithmic branch cut due to the One Pion
Exchange Potential. At $ k = \pm {\rm i} m_\pi $ a Two Pion Exchange
dilogarithmic branch cut sets in at about $E_{\rm LAB} \sim -40 {\rm
MeV} $, and so on. From this viewpoint any realistic description of
the phase shifts should be considered unreliable for energies beyond
the analyticity domain $ \sim 10 {\rm MeV} $. On the other hand, this
does not necessarily imply that the polylogarithmic corrections are
numerically large. So one may directly learn from the ``data'' how
important these corrections are in practical terms. There are two
possible ways to do this, either by using the SYM phase shifts given
in Ref.~\cite{Stoks:1993tb} or by directly evaluation of the scaled
M-matrix defined. In terms of the LAB
energy, $E_{\rm LAB}= k^2 / \mu_{np} $ the scaled $M$ matrix can be
represented as a constant, straight line and a parabola if we keep the
scattering lengths, effective ranges and curvature parameters. The
results of such a comparison for the scaled M-matrix is given in the
series of figures Fig.~(\ref{fig:M-matrix_j=0}),
(\ref{fig:M-matrix_j=1}), (\ref{fig:M-matrix_j=2}),
(\ref{fig:M-matrix_j=3}), (\ref{fig:M-matrix_j=4}) and
(\ref{fig:M-matrix_j=5}) for all partial waves with $ 0 \le j \le 5
$. 
In table~\ref{tab:table1bis} we present
the effective range parameters for each partial wave in units of
powers of $m_\pi /2 $, so that they correspond to the contribution of
different terms of the ERE at the maximum value of $k$ within the
domain of analyticity of $\hat {\bf M} (k) $.

On the light of these figures, two
conclusions may be drawn. In the first place, for a fixed LAB energy
as we increase the angular momentum, the {\it relative } error to the
full scaled M-matrix increases. This is expected since higher angular
momenta are more sensitive to the long distance physics, which is pion
dominated. In the limit of large angular momentum peripheral waves, the
low energy threshold parameters should be described solely in terms of
pion dynamics. The second observation is that the {\it absolute} error
decreases for increasing angular momentum as it corresponds for a
suppression of large scattering angles.

\subsection{Poles and Residues of the scattering matrix}
\label{sec:bound} 
The scattering amplitude has poles for negative energies~\footnote{We
assume that there are no degenerate poles. In the two channel case this
condition is automatically satisfied.}
\begin{eqnarray}
{\bf S}_{l,l'}^{sj} \to \frac{A_l^{sj} A_{l'}^{sj}}{\gamma + {\rm i} k }  
\label{eq:S-pole} 
\end{eqnarray} 
where $\gamma > 0$ corresponds to a bound state (first Riemann sheet
in $E$) and $\gamma < 0 $ (second Riemann sheet in $E$) to a virtual
state. The coefficients $A_l^{sj}$ correspond to the asymptotic bound
state wave function
\begin{eqnarray}
u_l^{sj} (r) \to A^{sj}_l \hat {\bf h}^{(+)} (i\gamma r)  
\end{eqnarray} 
In the $^1S_0$ channel one has $ \gamma=-\gamma_v $ and the asymptotic
wave function is
\begin{eqnarray}
u_{1S0} (r)  &\to&  A_{1S0} \, e^{\gamma_v r}
\end{eqnarray} 
In the case of the deuteron one has $\gamma=\gamma_d$ 
\begin{eqnarray}
u_{3S1} (r)  &\to&  A_{3S1} \, e^{-\gamma_d r} \\ 
u_{3D1} (r)  &\to&  A_{3D1} \, e^{-\gamma_d r} \left( 1 + \frac3{\gamma_d r} + 
 \frac3{(\gamma_d r)^2}  \right)  
\end{eqnarray} 
The ratio s to d wave is defined as
\begin{eqnarray}
\eta_d = \frac{A_{3D1}}{A_{3S1}} 
\end{eqnarray} 
The non-relativistic deuteron binding energy reads 
\begin{eqnarray}
B_d^{NR} = \frac{\gamma_d^2}{2 \mu_{np} }
\end{eqnarray} 
whereas the relativistic expression is given by 
\begin{eqnarray}
B_d^{R} = M_p + M_n -\sqrt{M_p^2- \gamma_d^2} - \sqrt{M_n^2- \gamma_d^2} 
\end{eqnarray} 
The calculation of threshold properties in the $^3S_1-^3D_1$ channel
from given deuteron properties is a standard procedure (see
e.g. Ref.~\cite{deSwart:1995ui}). Here we do just the opposite,
i.e. compute deuteron properties from threshold parameters within the
the effective range expansion. In practice this means solving the
equation
\begin{eqnarray}
{\rm Det } ({\bf \hat M} (k)^{-1} - i k {\bf D}^2) |_{k=i \gamma}  =0  
\label{eq:bound}
\end{eqnarray} 
using the LO,NLO and NNLO approximations. Residues are evaluated by
numerical integration using Cauchy's theorem.  The numerical results
for the smallest poles in the singlet $^1S_0$ and triplet
$^3S_1-^3D_1$ channels and their corresponding residues are presented
in Table~\ref{tab:table2}. As can be seen the expansion is convergent
but even to ${\rm N^4LO}$ in the amplitude is not sufficient to
reproduce the observables in a completely satisfactory way within
experimental uncertainties. The difference should in principle be
attributed to higher orders in the momentum expansion as well as other
effects.  In appendix~\ref{sec:app} we analyze these results further
on the light of an expansion around the limit of large s-wave
scattering lengths. Our conclusion is that such an expansion does not
improve on the description of the bound state observables. 

\begin{table*}
\caption{\label{tab:table2} Poles and residues of the scattering
amplitude for the NijmII and Reid93 (in brackets) potentials based on
the effective range expansion, Eq.~(\ref{eq:c-ere}). The data for
$B_{\rm d}$ are from \cite{le82}, for $\eta_{\rm d}$ from \cite{rod90}
and for $A_S$ from \cite{er83}.}
\begin{ruledtabular}
\begin{tabular}{|c|c|c|c|c|c|c|}
\hline 
& & & & & &\\ 
& LO   & NLO  & NNLO  & ${\rm N^3LO}$ & ${\rm N^4LO}$ & \\ 
& NijmII ( Reid93 ) & NijmII ( Reid93 ) & NijmII ( Reid93 ) 
& NijmII ( Reid93 ) & NijmII ( Reid93 ) 
& Exp. \\ \hline
\hline 
$\gamma_v ({\rm fm}^{-1}) $ &  -0.04215(-0.04213) & -0.04001(-0.03994) 
& -0.04001(-0.03994) & -0.04001(-0.03994) & -0.04001(-0.03994) &  \\
\hline \hline 
$\gamma_d ({\rm fm}^{-1}) $ & 0.18472(0.18438) & 0.22511(0.22498) 
& 0.23050(0.23045) & 0.23127(0.23120) & 0.23148(0.23141) & \\
$B^{NR}_d ({\rm MeV}) $ &  1.41197(1.40990) & 2.10146(2.09909) 
& 2.20345(2.20251) & 2.21807(2.21684) & 2.22215(2.22088) 
& 2.224575(9)\\
$B^{R}_d ({\rm MeV}) $ & 1.41250(1.41043) & 2.10263(2.10026)
&  2.20475(2.20380) & 2.21938(2.21815) & 2.22347(2.22219)  
& 2.224575(9)\\
$\eta_d $ & 0.01036(0.01033) & 0.02669(0.02659) 
& 0.02512(0.02502) & 0.02519(0.02510) & 0.02521(0.02512)
& 0.0256(4)\\
$A_{3S1}^d ( {\rm fm}^{-1/2} ) $ & 0.60732(0.60701) & 0.79351(0.79330) 
& 0.86620(0.86711) & 0.87805(0.87873) & 0.88219(0.88283) 
& 0.8846(9) \\
$A_{3D1}^d ( {\rm fm}^{-1/2} )$ & 0.00629(0.00627) & 0.02117(0.02110) 
& 0.02176(0.02170) & 0.02212(0.02205) & 0.02224(0.02218) &
\\
\end{tabular}
\end{ruledtabular}
\end{table*}

\section{Conclusions and Outlook} 
\label{sec:concl} 

In this paper we have extracted the low energy threshold parameters
for all $np$ partial waves up to states with total angular momentum $
j \le 5$ taking into account the coupled channel nature of the problem
for realistic NN potentials. Our description entails up to order
${\cal O}(k^8) $ in the CM momentum of the full np scattering
amplitude. An adequate framework is to use the effective range
expansion of the scaled coupled channel M-matrix, where the kinematic
centrifugal factors have been factorized out. These low energy
parameters comprise the scattering length matrix ${\bf a}$, the
effective range matrix ${\bf r}$, the curvature matrix ${\bf v_2}$ and
higher like ${\bf v_3}$ and ${\bf v_4} $ and are relevant for a shape
independent description of the scattering data in terms of the scaled
M-matrix in a region of analyticity in the complex energy plane around
the origin which radius extends up to the left partial wave
logarithmic cut generated by One Pion Exchange intermediate
states. The practical determination of these parameters from the
solutions of the Schr\"odinger equation may be cumbersome, so we have
found extremely convenient to use the variable S-matrix formalism. The
low energy threshold parameters can be directly determined from the
asymptotic solution of a set of coupled non linear differential
equations which correspond to an adiabatic switching on of the NN
potential using trivial initial conditions. Finally, we have found
that the coupled channel effective range expansion works well within
the expected region of analyticity, namely $ k \le m_\pi /2 $ and in
fact a clear trend to convergence is observed. However, beyond the
region of analyticity we do not expect the low energy expansion to be
realistic regardless on how many terms are included in the
expansion. Actually, the analyticity domain of the amplitude can be
enlarged if the proper left cut singularities, corresponding to
OPE,TPE, etc. are implemented. For such a program the method presented
in our previous work~\cite{Valderrama:2003np,Valderrama:2004nb} looks
particularly promising. Finally, one problem with the use of high
quality potentials has to do with the determination of errors on the
potentials and hence on the low energy parameters. In this paper we
have used two such potentials to assess those errors, but it would be
rather interesting to make the error analysis directly based on the
coupled channel effective range expansion.

\begin{acknowledgments}

We thank J. Nieves for reading the manuscript. 
This work is supported in part by funds provided by the Spanish DGI
with grant no. BMF2002-03218, Junta de Andaluc\'{\i}a grant no. FM-225
and EURIDICE grant number HPRN-CT-2003-00311.

\end{acknowledgments}

\appendix 
\section{Difficulties in extracting the low energy parameters from a fit} 
\label{sec:app3} 

At first sight one might think that the effective range parameters
could be determined directly from a fit to the Nijmegen data
base~\cite{Stoks:1993tb} and avoid the use of the corresponding
potentials. In this appendix we want to elaborate on the problems we
have encountered while fitting that data base within a generalized
coupled channel effective range expansion of
Eq.~(\ref{eq:c-ere}). Unfortunately, this data base does not provide
error estimates for their phase shifts (although 8 significant digits
are given), nor the typical energy resolution where these data should
be trusted, so some compromise must be made.

\begin{figure*}
\begin{center}
\epsfig{figure=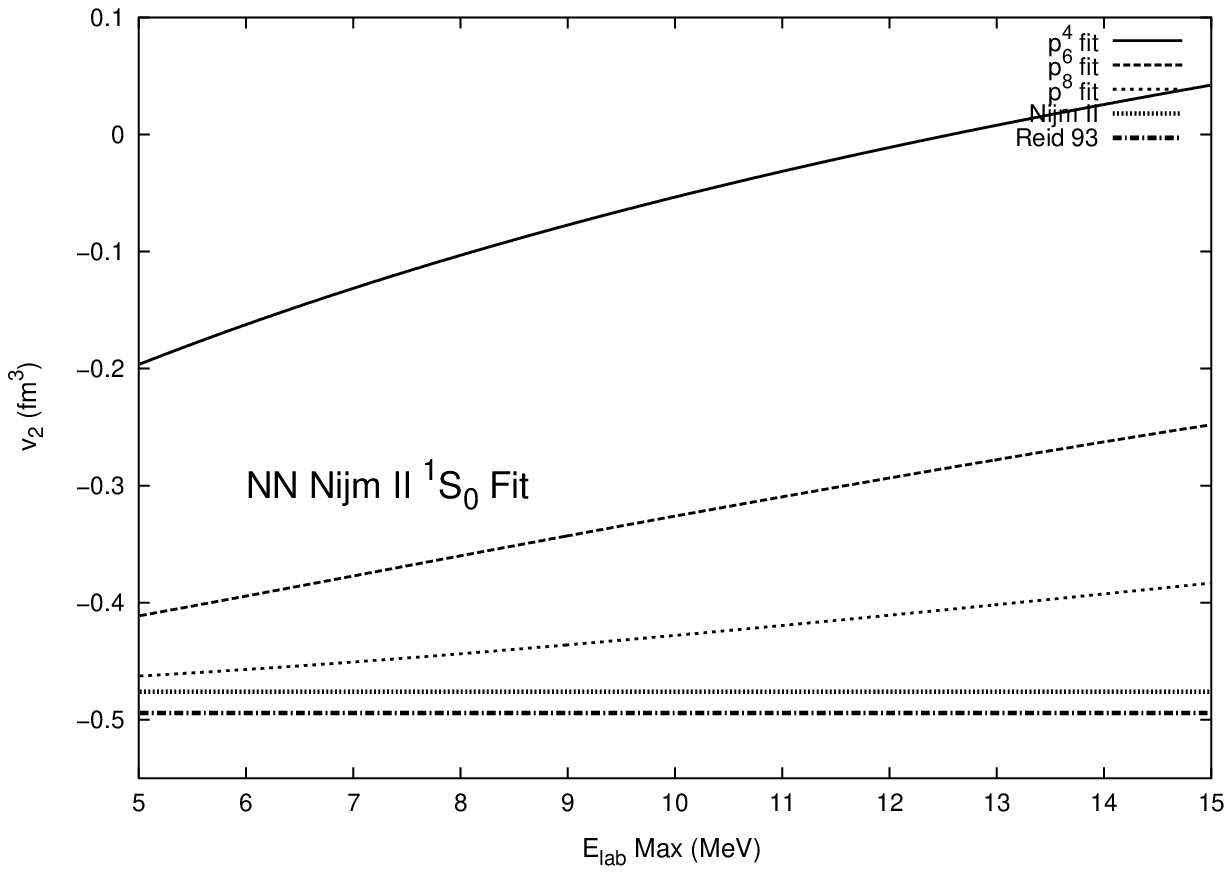,height=8cm,width=8cm} 
\epsfig{figure=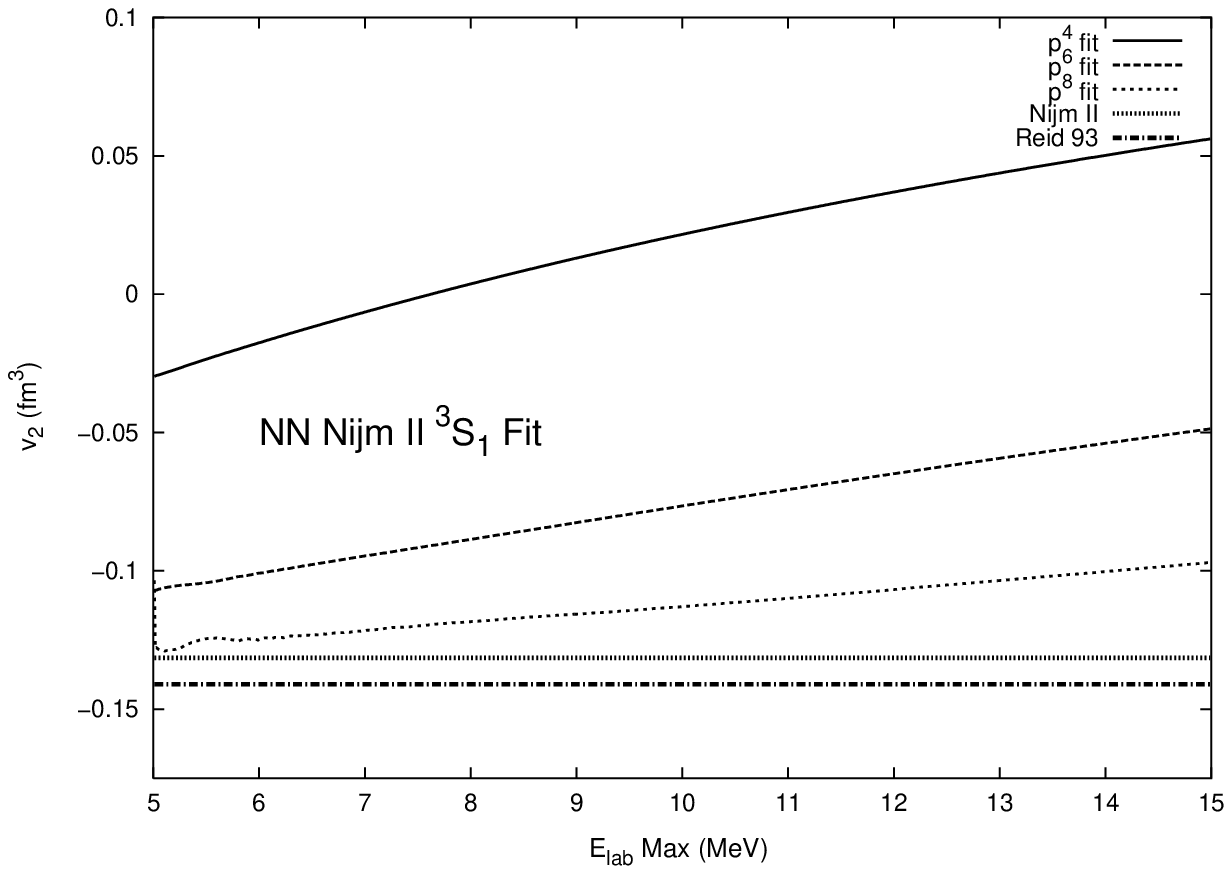,height=8cm,width=8cm} 
\end{center}
\caption{The $v_2$ parameter for the $^1S_0$ (left) and the $^3S_1$
(right) channels determined from a fit to the low energy date of the
NN data base \cite{Stoks:1993tb} (see Eq.~(\ref{eq:chi2})) and main
text, as a function of the maximal LAB-energy considered in the fit.
$p^n$ means a fit including up to ${\cal O} (k^n) $ terms in the
effective range expansion Eq.~\ref{eq:era_fit}. ``Database'' means a
fit to the average value of the corresponding scaled
$M$-matrix. ``Reid93'' and ``NijmII'' means a fit to only this
data. The values we obtained by integrating the Eqs.~(\ref{eq:var})
NijmII and Reid93 potentials. }
\label{fig:lecs_fit}
\end{figure*}

We use the NN-data and define the $\chi^2$ as
\begin{eqnarray}
\chi^2 = \sum_{i=1}^N \left( \frac{\hat M_{\rm ER} - \bar M_{\rm NN}}{\Delta
M_{\rm Nm} }\right)^2   \frac{M}{4k} 
\label{eq:chi2}
\end{eqnarray} 
where we take $ \Delta E_{\rm LAB } =0.01 {\rm MeV} $ , and $ \bar
M_{\rm NN} $ and $ \Delta M_{\rm NN} $ are the mean value and the
standard deviation of the six potentials listed in the NN-data
base~\cite{Stoks:1993tb}, which can be taken as independent
uncorrelated primary data.  The factor $M/(4k) $ is the Jacobian of
the transformation between the Lab-energy and the C.M. momentum, $
E_{\rm LAB} = 2 k^2 /M $, and would correspond to make an equidistant
sampling in $p$, in the limit $ \Delta E_{\rm LAB} \to 0 $ (this is
why we take a small energy spacing). This weight factor is introduced
in order to enhance the region at low momenta. On the other hand, very
low momenta must be excluded since the resulting mean value $M$-matrix
is incompatible within the attributed errors with the expected
theoretical behaviour, Eq.~(\ref{eq:c-ere}), so we take $E_{\rm LAB}
\ge 0.5 {\rm MeV} $. This is partly due to the poor accuracy of the
data at low energies; the calculation of the scaled M-matrix requires
increasing accuracy at low energies. Also, the fit goes up to $E_{\rm
LAB} \le 10 {\rm MeV} $, which corresponds to a C.M. momentum about $
k = m_\pi /2 $ where we expect the finite polynomial of the scaled
$M-matrix $ to truly represent an analytical function within the
convergence radius up to the branch cut singularity located at $ k =
\pm {\rm i} m_\pi /2 $. The form of the fitting function is
\begin{eqnarray}
\hat M_{\rm ER}= v_0+ v_1  k^2 + v_2 k^4+ v_3 k^6 + v_4 k^8
+ \dots
\label{eq:era_fit} 
\end{eqnarray} 
In Fig.~\ref{fig:lecs_fit} we show as an illustration the $v_2$
parameter determined from a fit to the low energy region of the NN
data base \cite{Stoks:1993tb} as a function of the maximal LAB-energy
considered in the fit. As we see, instead of a plateau within some
energy window, we observe an ever changing value. We observe no
stability depending on the number of terms considered in
Eq.~(\ref{eq:era_fit}) either. For comparison we also plot the values we
obtained by integrating the Eqs.~(\ref{eq:var}) with the NijmII potential in
Sect.~\ref{sec:v_threshold}, which where quite stable numerically. As we see,
the values obtained from the fit, in the chosen energy window are
hardly compatible. The deceptive features extend to other channels,
and non diagonal low energy threshold parameters such as the matrix
elements of ${\bf a} $ and ${\bf r}$.

Finally, we have also tried, with no success, other methods for the
determination of the low energy threshold parameters, like evaluation
of derivatives within several algorithms. The reason for the failure
has to do with round-off errors generated by the relatively small
number of digits provided in the NN database. Actually, at very low
energies these round-off errors make the construction of the scaled M
matrix itself rather unstable numerically, since large centrifugal
factors $1/k^l $ are involved.

\section{The limit of large scattering lengths} 
\label{sec:app} 

In this appendix we discuss some interesting aspects of the limit of
large scattering lengths. For a $s-$wave eigen phase shift,
i.e. $^1S_0$ and $^3S_1$ eigen channels, the effective range expansion
can be written as
\begin{eqnarray}
k \cot \delta = - \frac1\alpha_0 + \frac12 r_0 k^2 + v_2 k^4 + v_3 k^6
+ v_4 k^8 + \dots   
\end{eqnarray} 
The poles in the S-matrix are given by the solutions of the equation 
\begin{eqnarray}
f_0 ( i \gamma )^{-1} = k \cot \delta - {\rm i} k \Big|_{k={\rm i} \gamma }
=0 
\label{eq:gamma_def} 
\end{eqnarray} 
For a given order of the truncated effective range expansion the
previous equation reduces to an algebraic equation with real
coefficients and it has at least one real solution, which can be taken
as the analytical continuation of the lower order ones by taking the
limit of the effective range parameters $r_0, v_2 , \dots \to 0$, or
alternatively as the limit $\alpha_0 \to \infty $. All other roots
either wander to infinity or accumulate to build, in the limit of
infinitely many terms of the expansion, a branch cut.  Expanding into
powers of $1/\alpha_0$ one gets
\begin{eqnarray}
\gamma &=& \frac{1}{\alpha_0} + \frac{r_0}{2 \alpha_0^2} +
 \frac{r_0^2}{2\alpha_0^3} + \frac{\left(\frac58 r_0^3 -
 v_2\right)}{\alpha_0^4} \nonumber \\ &+& \frac{\left(\frac78 r_0^4 -
 3 r_0 v_2 \right)}{\alpha_0^5} + \frac{\left(\frac{21}{16} r_0^5 - 7
 r_0^2 v_2 + v_3 \right)}{\alpha_0^6} \nonumber \\ &+&
 \frac{\left(\frac{33}{16} r_0^6 - 15 r_0^3 v_2 + 4 v_2^2 + 4 r_0 v_3
 \right)}{\alpha_0^7}+ \dots 
\label{eq:gamma_exp}
\end{eqnarray} 
The neglected terms contain the parameters $v_4, \dots $. Within the
same approximation the residue for the pole of the S-matrix in this
(eigen)channel defined in analogy to Eq.~(\ref{eq:S-pole}), $ S_0^2
\to A_0^2 / ( \gamma_d + {\rm i} k  ) $ becomes 
\begin{eqnarray}
A_0^2 &=& \frac2{\alpha_0} + \frac{3 r_0}{\alpha_0^2}+ \frac{5
r_0^2}{\alpha_0^3}+ \frac{\frac{35}4 r_0^3 -10 v_2 }{\alpha_0^4}
\nonumber \\ &+& \frac{ \frac{63}4 r_0^4 - 42 r_0 v_2 }{\alpha_0^5}+
\frac{ \frac{231}8 r_0^5 -126 v_2 r_0^2 + 2 v_3}{\alpha_0^6} \nonumber
\\ &+& \frac{\frac{429}8 - 330 v_2 r_0^3 + 12 v_3 r_0 + 72
v_2^2}{\alpha_0^7} + \dots
\label{eq:res_exp}
\end{eqnarray} 
The expansions Eq.~(\ref{eq:gamma_exp}) and Eq.~(\ref{eq:res_exp}) are
convergent provided $\alpha_0 $ is within the domain of analyticity of
the exact solution of Eq.~(\ref{eq:gamma_def}). Otherwise the
expansion is asymptotic, i.e. we can use it to evaluate numerically
the value of $\gamma $ in the limit $\alpha_0 \to \infty $, up to a
given order where the remainder starts increasing basically due to the
presence of large factorials. Nevertheless one can use the expansion
with a given error estimate. For instance, at NLO the domain of
analyticity is given by the condition of a vanishing discriminant of a
second order algebraic equation yielding $ \alpha_0 = 2 r_0 = 3.50652$
for the analyticity inner boundary. The numerical values for
$\alpha_0$ and $r_0$ lie within the boundary and one can use the
series expansion to any order to evaluate $\gamma$ with increasing
accuracy. In the NNLO case the analyticity boundary is given by an
analogous discriminant condition. Numerically we find the inner
boundaries for $\alpha_0 $ located at lower points $\alpha_0 =
3.444,-0.777,-0.133$ for the NijmII potential.

\begin{table*}
\caption{\label{tab:table3} Poles and residues of the scattering
amplitude for the NijmII potential based on the effective range
expansion, in the large scattering length limit $\alpha \to \infty $
to ${\cal O} (1 /\alpha^5) $ with $\alpha = \alpha_{^1S_0}$ in the
$^1S_0$ channel, $\alpha = \alpha_{3S1}$ in the $^3S_1-^3D_1$ 
channel ($^3S_1$ eigen channel),
and $\alpha= \bar \alpha_{3S1}$ in the ``bar'' $^3S_1-^3D_1$ channel
(barred quantities). This calculation is complete to ${\rm N^4 LO}$
(see main text).}
\begin{ruledtabular}
\begin{tabular}{|c|c|c|c|c|c|}
\hline  & & & & &  \\ 
&${\cal O}( 1/\alpha)$ & ${\cal O}( 1/\alpha^2)$
&${\cal O}( 1/\alpha^3 )$ & ${\cal O}( 1/\alpha^4) $ 
&${\cal O}( 1/\alpha^5)$ \\
& NijmII ( Reid93 ) & NijmII ( Reid93 ) & NijmII ( Reid93 ) 
& NijmII ( Reid93 ) & NijmII ( Reid93 ) \\
\hline
\hline 
$\gamma_v ({\rm fm}^{-1}) $ & -0.04215(-0.4213) & -0.03977(-0.03969) 
& -0.04004(-0.03997) & -0.4000(-0.03997) & -0.4001(-0.03994) \\
\hline \hline 
${\gamma_d} ({\rm fm}^{-1}) $ & 0.18454(0.18440) & 0.21439(0.21425) 
& 0.22405(0.22391) & 0.22791(0.22779) & 0.22962(0.22952) \\
$B^{NR}_d ({\rm MeV}) $ & 0.0(0.0) & 1.41227(1.41020)
& 1.86922(1.86673) &  2.05403(2.05146) & 2.13694(2.13461) \\
$B^{R}_d ({\rm MeV}) $ & 0.0(0.0) & 1.41227(1.41020)
& 1.86922(1.86673) & 2.05456(2.05199) & 2.13782(2.13549) \\
$\eta_d $ & 0.0(0.0) & 0.0(0.0) & 0.01035(0.01032) & 0.01218(0.01212) & 0.01671(0.01663) \\
$({ A}_{3S1}^d)^2 ( {\rm fm}^{-1} ) $ & 0.36908(0.36880) 
& 0.54820(0.54789) & 0.64479(0.64452) & 0.69896(0.69889) & 0.73010(0.73027) \\
\hline \hline
$\bar{\gamma_d} ({\rm fm}^{-1}) $ & 0.18454(0.18440) & 0.21576(0.21558)  
& 0.23050(0.23037) & 0.23209(0.23197) & 0.23413(0.23404) \\
$\bar{B}^{NR}_d ({\rm MeV}) $ & 0.0(0.0) & 1.41227(1.41020)  
& 1.89010(1.88689) & 2.15612(2.15363) & 2.21874(2.21629) \\
$\bar{B}^{R}_d ({\rm MeV}) $ & 0.0(0.0) & 1.41227(1.41020)
& 1.89010(1.88689) & 2.15665(2.15416) & 2.21963(2.21718) \\
$\bar{\eta}_d $ & 0.0(0.0) & 0.0(0.0) & 0.01035(0.01032) & 0.01233(0.01227) 
& 0.01741(0.01733) \\
$(\bar{ A}_{3S1}^d)^2 ( {\rm fm}^{-1} ) $ & 0.36907(0.36880) 
& 0.55639(0.55584) 
& 0.68707(0.68679) & 0.74308(0.74290) & 0.78694(0.78718) \\
\end{tabular}
\end{ruledtabular}
\end{table*}

\begin{table*}
\caption{\label{tab:table3bis} Same as table \ref{tab:table3} 
but for higher orders.}
\begin{ruledtabular}
\begin{tabular}{|c|c|c|c|c|c|}
\hline  & & & & &  \\ 
&${\cal O}( 1/\alpha^6)$ & ${\cal O}( 1/\alpha^7)$
&${\cal O}( 1/\alpha^8 )$ & ${\cal O}( 1/\alpha^9) $ 
&$ {\rm N^4LO} $ \\
& NijmII ( Reid93 ) & NijmII ( Reid93 ) & NijmII ( Reid93 ) 
& NijmII ( Reid93 ) & NijmII ( Reid93 ) \\
\hline
\hline 
$\gamma_v ({\rm fm}^{-1}) $ & -0.04001(-0.03994) & -0.04001(-0.03994)
& -0.04001(-0.03994) & -0.04001(-0.03994) & -0.04001(-0.3994) \\
\hline \hline 
${\gamma_d} ({\rm fm}^{-1}) $ & 0.23047(0.23038) & 0.23091(0.23083) 
& 0.23116(0.23109) & 0.23131(0.23123) & 0.23148(0.23141)\\
$B^{NR}_d ({\rm MeV}) $ & 2.17666(2.17462) & 2.19696(2.19515)  
& 2.20784(2.20620) & 2.21396(2.21245) & 2.22215(2.22088) \\
$B^{R}_d ({\rm MeV}) $ & 2.17773(2.17568) & 2.19814(2.19632) 
& 2.20908(2.20744) & 2.21523(2.21371) & 2.22347(2.22219)\\
$\eta_d $ &  0.01848(0.01838) & 0.02053(0.02042) & 0.02167(0.02156) 
& 0.01953(0.01942) & 0.02521(0.02512) \\
$({ A}_{3S1}^d)^2 ( {\rm fm}^{-1} ) $ & 0.74867(0.74907) 
& 0.76000(0.76059) & 0.76717(0.76790) & 0.77181(0.77265) & 0.77826(0.77939)\\
\hline \hline
$\bar{\gamma_d} ({\rm fm}^{-1}) $ & 0.23344(0.23336) & 0.23356(0.23349) 
&  0.23289(0.23282) &  0.23272(0.23266) & 0.23148(0.23141)\\
$\bar{B}^{NR}_d ({\rm MeV}) $ & 2.26308(2.26123) & 2.25980(2.25807)  
& 2.26231(2.26093) & 2.25179(2.25044) & 2.22215(2.22088) \\
$\bar{B}^{R}_d ({\rm MeV}) $ & 2.26423(2.26238) & 2.26106(2.25934)
&  2.26365(2.26226) & 2.25314(2.25179) & 2.22347(2.22219)\\
$\bar{\eta}_d $ & 0.01895(0.01886) & 0.02149(0.02140) 
& 0.02238(0.02228)  & 0.01979(0.01969) & 0.02521(0.02512) \\
$(\bar{ A}_{3S1}^d)^2 ( {\rm fm}^{-1} ) $ & 0.79951(0.79996)
& 0.81118(0.81198) & 0.80941(0.81033) & 0.80990(0.81102) 
& 0.77826(0.77939)\\
\end{tabular}
\end{ruledtabular}
\end{table*}

If we instead make the large scattering length expansion in the
determinant bound state condition, Eq.~(\ref{eq:bound}), for a
truncated scaled M-matrix we get an algebraic equation in the deuteron
pole $ k= {\rm i} \gamma $ which order is determined by the level of
approximation. In the SYM parameterization the LO, NLO, NNLO, NNNLO
effective range parameters start contributing at order $\gamma$,
$\gamma^2$, $\gamma^4$ and $\gamma^6$ respectively and also contribute
to a maximal order $\gamma^6$, $\gamma^7$, $\gamma^9$ and
$\gamma^{12}$ respectively. This means that the coefficients of
$\gamma$ in an expansion in $1/ \bar \alpha_{3S1} $ are complete to $
{\cal} O(\bar \alpha_{3S1}^{-1} )$ at LO, $ {\cal} O(\bar
\alpha_{3S1}^{-3} )$ at NLO, and $ {\cal} O(\bar \alpha_{3S1}^{-5}) $
at NNLO. This can also be seen by applying the large scattering length
expansion of Eq.~\ref{eq:gamma_exp} to the eigen phase shift,
Eq.~(\ref{eq:keigen}), with the identifications (\ref{eq:sym-eigen}),
yielding for the pole
\begin{eqnarray}
\gamma_d &=& \frac{1}{ \bar \alpha_{3S1}}+ \frac12 \frac{\bar
r_{3S1}}{\bar \alpha_{3S1}^2}+\frac12 \left( \bar r_{3S1}^2 + 2 \bar
\alpha_{E1} \bar r_{E1} \right) \bar \alpha_{3S1}^{-3} \nonumber \\
&+& \frac14 \Big( 2 \bar \alpha_{E1}^2 \bar r_{3D1} + \frac52 \bar
r_{3S1}^3 +8 \bar \alpha_{E1} \bar r_{3S1} \bar r_{E1}
 \nonumber \\  && - \bar \alpha_{3D1} \bar r_{E1} - 4
\bar v_{3S1} \Big) \bar \alpha_{3S1}^{-4}+ \dots 
\end{eqnarray} 
We also get for the residues 
\begin{eqnarray}
A_{3S1}^d A_{3D1}^d =  \frac{2 \bar \alpha_{E1}}{\bar \alpha_{3S1}^4}
- \frac{ \bar \alpha_{3D1} \bar r_{E1} - 5 \bar \alpha_{E1} \bar
r_{3S1}^2 }{\bar \alpha_{3S1}^2} + \dots   
\end{eqnarray} 
Finally, for the mixing coefficient $ \eta_d $ we get 
\begin{eqnarray} 
\eta_d &=& \frac{\bar \alpha_{E1}}{\bar \alpha_{3S1}^3} + \frac{ 2
\bar \alpha_{E1} \bar r_{3S1} - \bar \alpha_{3D1} \bar r_{E1}}{2 \bar
\alpha_{3S1}^4} + \dots 
\label{eq:eta_pert} 
\end{eqnarray} 
and so on.  Note that the leading contributions in the inverse
scattering length behave differently for any observable.  According to
Eq.~(\ref{eq:sym-eigen}), making $\alpha_{3S1} $ large makes also $
\bar \alpha_{3S1} $ large. Thus one may study separately the
convergence of observables in both cases. The numerical values for the
pole position and corresponding residua evaluated in the inverse
scattering length expansion are presented in Tables~\ref{tab:table3}
for the expansion up to order $1/\alpha^5 $ and $1/\bar
\alpha^5$. This order involves for pole position only up to ${\rm N^4
LO }$ in the amplitude (for the binding energy and mixing parameter
the situation is somewhat different, see
e.g. Eq.~(\ref{eq:eta_pert})). As we have already discussed the
converse is not true, and ${\rm N^4LO} $ contains a series of higher
power corrections in the inverse scattering length. We present those
corresponding only to ${\rm N^4LO}$ in Table~\ref{tab:table3bis} up to
ninth order in the inverse scattering length. As we see, and one could
have anticipated, the convergence for the virtual state pole is faster
than for the deuteron bound state. Actually, we see that the expansion
in the inverse scattering length for bound state properties is less
convergent than the ERE pursued in Sect.~\ref{sec:bound}.


\end{document}